\documentclass[a4paper,11pt]{article}

\usepackage{pos}

\title{
  Use of Schwinger-Dyson equation in constructing an approximate trivializing map
}
\ShortTitle{Use of Schwinger-Dyson equation in constructing an approximate trivializing map}

\author[a,b]{Peter Boyle}
\author[b,c]{Taku Izubuchi}
\author[d]{Luchang Jin}
\author[b]{Chulwoo Jung}
\author[e]{Christoph Lehner}
\author*[c]{Nobuyuki Matsumoto}
\author[f]{Akio Tomiya}

\affiliation[a]{
  School of Physics and Astronomy, University of Edinburgh, Edinburgh EH9 3FD, United Kingdom
}
\affiliation[b]{
  Physics Department, Brookhaven National Laboratory, Upton, NY 11973, USA
}
\affiliation[c]{
  RIKEN/BNL Research center, Brookhaven National Laboratory, 
  Upton, NY 11973, USA
}
\affiliation[d]{
  Physics Department, University of Connecticut, Storrs, CT 06269, USA
}
\affiliation[e]{
  Universit \"{a}t Regensburg, Fakult\"{a}t f\"{u}r Physik, 93040 Regensburg, Germany
}
\affiliation[f]{
  Faculty of Technology and Science, International Professional University of Technology,
  Osaka, Japan
}

\emailAdd{nobuyuki.matsumoto@riken.jp}

\abstract{
  We construct an approximate trivializing map
  by using a Schwinger-Dyson equation.
  The advantage of this method
  is that: (1) The basis for the flow kernel can be chosen arbitrarily by hand.
  (2) It can be applied to the general action of interest.
  (3) The coefficients in the kernel are determined by lattice estimates of the observables,
  which does not require analytic calculations beforehand.
  We perform the HMC with the effective action obtained by the Schwinger-Dyson method,
  and show that we can have better control of the effective action than
  the known $t$-expansion construction.
  However, the algorithmic overhead is still large and overwhelming the gain
  though faster decorrelation is observed for long-range observables in some cases.
  This contribution reports the preliminary results of this attempt.
}

\FullConference{%
  The 39th International Symposium on Lattice Field Theory (Lattice2022),\\
  8-13 August, 2022 \\
  Bonn, Germany 
}

\begin{document}
\maketitle

\section{Introduction}
\label{sec:introduction}

In recent years, lattice QCD has been taking an important and crucial role in the
precision test of the standard model,
as represented by the calculation of QCD contributions to the muon g-2
(see, e.g., \cite{Lehner:2022}).
Empowered by the recent hardware and experimental developments,
target precision is becoming in a notable order,
for which calculation on fine lattices has become an urgent demand.
However, as we reach the continuum limit, we face the infamous critical slowing down,
often characterized by long autocorrelation of the topological charge.
Such long autocorrelations make the calculation at fine lattices
inefficient adding extreme computational cost to the simple lattice volume scaling.

There have been many attempts to overcome the critical slowing down
\cite{Parisi:1984,Batrouni:1985jn,Davies:1987vs,Katz:1987ti,
  Davies:1989vh,Luscher:2009eq,Luscher:2011kk,Albergo:2019eim,Foreman:2021ixr,
  Albandea:2021lvl,Foreman:2021ljl,Nguyen:2021zgx},
and recent studies are presented thoroughly at the conference \cite{
  Urban:2022,Cellini:2022,Lopez:2022,Shanahan:2022,
  Rossney:2022,Albandea:2022,Komijani:2022,Francis:2022,
  Eichhorn:2022,Rouenhoff:2022,Swaim:2022,Bonanno:2022,
  Kara:2022,DAngelo:2022,Barros:2022,Finkenrath:2022}.
We in this work concentrate on developing the idea of
the {\it trivializing map}
proposed by L\"uscher \cite{Luscher:2009eq}
(cf. the Nicolai map \cite{Nicolai:1980jc}),
which maps a finite $\beta$ theory to the $\beta$ = 0 theory.
Evidently, if we manage to construct such a map,
one can obtain configurations on a fine lattice
from the configurations of purely random $SU(3)$ fields.

In \cite{Luscher:2009eq}, L\"uscher proved the existence of the map
in the form of a gradient flow,
and gave a way to construct the flow
as a $t$-expansion ($t$:~flow time)
as we review in section~\ref{sec:triv_map}.
To the leading order in $t$,
this flow corresponds to the Wilson flow
for the plaquette action.
Following this work, Engel and Schaefer \cite{Engel:2011re}
tested the method
by implementing the leading order gradient flow
in a $CP^{N-1}$ model, whose result asserted its performance to be
rather negative observing the scaling towards the continuum limit.

The aim of this work is to
give an alternative way to construct an approximate trivializing map.
In our method, the coefficients in the flow kernel are determined
from lattice estimates of Wilson loops by using a Schwinger-Dyson equation.
This method is versatile in the sense that:
(1) The basis for the flow kernel can be chosen by hand.
(2) It can be applied to the general action of interest
without an analytical calculation.
Here, the truncation effects and goodness of the flow
can be evaluated by a force norm.
We perform the HMC \cite{Duane:1987de}
with the effective action
obtained by the Schwinger-Dyson method.
The configuration generation part of the algorithm is
the same as \cite{Luscher:2009eq}.
We apply our method to Wilson and DBW2 actions and show that
we can have better control of the effective action than the $t$-expansion.
Furthermore, in some cases,
faster decorrelation (in Monte Carlo steps)
is observed for long-ranged observables by adding
the extended shapes such as rectangle and chair to the flow;
however, the algorithmic overhead is still large and overwhelming the gain.
This contribution therefore
reports the preliminary results in this direction.

The rest of this paper is organized as follows.
Section~\ref{sec:triv_map} is a review of the trivializing map
including the $t$-expansion.
In section~\ref{sec:schwinger_dyson},
we first introduce a Schwinger-Dyson equation
that determines the coefficients of Wilson loops in the effective action.
We then describe its use in constructing an approximate
trivializing map.
Section~\ref{sec:results} describes the
results focusing on the norm measuring the closeness
of the effective action to the target action
and the autocorrelation of the smeared energy density.
Section~\ref{sec:summary} is devoted for discussion.

 \section{Trivializing map}
\label{sec:triv_map}

\subsection{Review of L\"uscher's construction}
\label{sec:luscher_construction}

Following L\"uscher \cite{Luscher:2009eq},
we consider a field transformation: $\mathcal{F}: V \mapsto U$,
by which the original action $S(U)$ of interest will be mapped to
the effective action $S_{\rm eff}(V)$ according to the formula:
\begin{align}
  \int (dU)\, e^{-S(U)}
  &= \int (dV)\, {\rm det}\mathcal{F}_*(V) \, e^{-S(\mathcal{F}(V))} \nonumber\\
  &\equiv \int (dV)\, e^{-S_{\rm eff}(V)},
\end{align}
i.e.,
\begin{align}
  S_{\rm eff}(V) \equiv S(\mathcal{F}(V)) - \ln \det \mathcal{F}_*(V).
\end{align}
The Jacobian matrix $\mathcal{F}_*(V)$
can be defined by the
local parameterization $\theta_{x,\mu}^a$ of the field space:
\begin{align}
  e^{\theta_{x,\mu}^a T^a} U_{x,\mu},
\end{align}
where $T^a$ are the ${\mathfrak su}(3)$ generators
normalized as ${\rm tr}\, T^aT^b = -(1/2)\delta^{ab}$.
Note that the Haar measure can then be written as:
\begin{align}
  (dU) = {\rm const.} \, \prod_{x,\mu,a} d\theta_{x,\mu}^a.
\end{align}
The derivative $\partial_{x,\mu}^a \equiv \partial_{\theta_{x,\mu}^a}|_{\theta=0}$
acts as 
\begin{align}
  \partial_{x,\mu}^a U_{y,\nu} = \delta_{x,y}\delta_{\mu,\nu} T^a U_{x,\mu}. 
\end{align}
Writing the indices in a short way as $A\equiv(x,\mu,a)$,
the Jacobian matrix $\mathcal{F}_*(V) \equiv (\mathcal{F}^{AB}_*(V))$
relates the one-forms of the $U$-space and the $V$-space:
\begin{align}
  d\theta^A_{U} \equiv \mathcal{F}^{AB}_*(V) d\theta^B_{V},
\end{align}
where the subscript is supplied to distinguish the coordinates of
the two field spaces.

L\"uscher obtained the {\it trivializing map}
by considering a flow ${\mathcal{F}_t}: V \mapsto U_t \equiv {\mathcal{F}}_t(V)$
and by demanding the effective action at flow time $t$ to be:
\begin{align}
  S_{{\rm eff}, t}(V)
  \stackrel{*}{=}
  (1-t) S({\mathcal{F}_t(V)}),
  \label{eq:demand}
\end{align}
up to an irrelevant $t$-dependent constant
which we ignore hereafter.
The symbol $\stackrel{*}{=}$
means it is a relation we require
so that the $V$-space action becomes trivial at $t=1$.
We divide the trivializing map ${\mathcal{F}}_{t=1}$
into infinitesimal steps:
\begin{align}
  \mathcal{F}_{t=1} =
  \mathcal{F}_{(n-1)\epsilon,\epsilon} \circ \mathcal{F}_{(n-2)\epsilon,\epsilon} \circ
  \cdots \circ \mathcal{F}_{0, \epsilon},
  \label{eq:composition_ordinary}
\end{align}
where $1=n\epsilon$, and use the gradient flow ansatz:
\begin{align}
  \mathcal{F}_{t, \epsilon}(U)_{x,\mu}
  = e^{- \epsilon T^a \partial_{x,\mu}^a \tilde{S}_t(U)} U_{x,\mu}.
  \label{eq:ansatz}
\end{align}
The derivative acts on the direct argument of the function.\footnote{
  Therefore, the expression $\partial_{x,\mu}^a \tilde{S}_t(U)$
  can be thought of as a vector function of $U$:
  $\partial_{x,\mu}^a \tilde{S}_t(U) = (\partial_{x,\mu}^a \tilde{S}_t)(U)$.
  We use this convention throughout this contribution.
}
Using that, for $U' \equiv \mathcal{F}_{t, \epsilon}(U)$,
$d=d\theta_{U'}^A \partial_{U'}^A=d\theta_U^A \partial_U^A$ and
\begin{align}
  {d\theta_{U'}}_{x,\mu}^a
  = -2 {\rm tr}[T^a dU'_{x,\mu} U_{x,\mu}^{\prime -1}],
\end{align}
one finds for an infinitesimal $\varepsilon$,
\begin{align}
  { \mathcal{F}_{t, \epsilon, *}}\, ^a_{x,\mu}\vert^b_{y,\nu}
  = \delta_{x,y}\delta_{\mu,\nu}\delta^{ab}
  -\epsilon \partial_{x,\mu}^a\partial_{y,\nu}^b\tilde{S}_t
  -\epsilon \delta_{x,y}\delta_{\mu,\nu} f^{abc}\partial_{x,\mu}^c\tilde{S}_t,
\end{align}
where $f^{abc} \equiv -2{\rm tr}\big(T^a[T^b,T^c]\big)$
is the totally antisymmetric tensor.
Therefore,
\begin{align}
  \ln {\rm det}\, { \mathcal{F}_{t, \epsilon, *}} = - \epsilon (\partial^A)^2{\tilde S}_t.
\end{align}
We note that the ansatz~\eqref{eq:ansatz}
is equivalent to:
\begin{align}
  \dot{U}_{t, x, \mu} = \frac{d}{dt}\mathcal{F}_{t}(V)_{x,\mu} = - T^a \partial_{x,\mu}^a \tilde{S}_t(U_t) \, U_{t,x,\mu}.
  \label{eq:config_space_flow}
\end{align}
Rewriting eq.~\eqref{eq:demand} as
\begin{align}
  \ln \det \mathcal{F}_{t,*}(V)
  \stackrel{*}{=}
  - t S({\mathcal{F}_t(V)}),
\end{align}
and by taking the $t$-derivative,
we obtain the functional equation:
\begin{align}
  [-(\partial^A)^2 + t (\partial^A S) \partial^A]\tilde{S}_t \stackrel{*}{=} S.
  \label{eq:diff_eq}
\end{align}
Since the differential operator
${\mathcal L}_t\equiv -(\partial^A)^2 + t (\partial^A S) \partial^A$
is elliptic and
there is an inner product that makes ${\mathcal L}_t$ a symmetric operator,
it is invertible for finite lattice up to the constant function,
which is the zero-mode \cite{Luscher:2009eq}.
This fact suggests that the trivializing map formally exists.

\begin{figure}[htb]
  \centering
  \includegraphics[width=100mm]{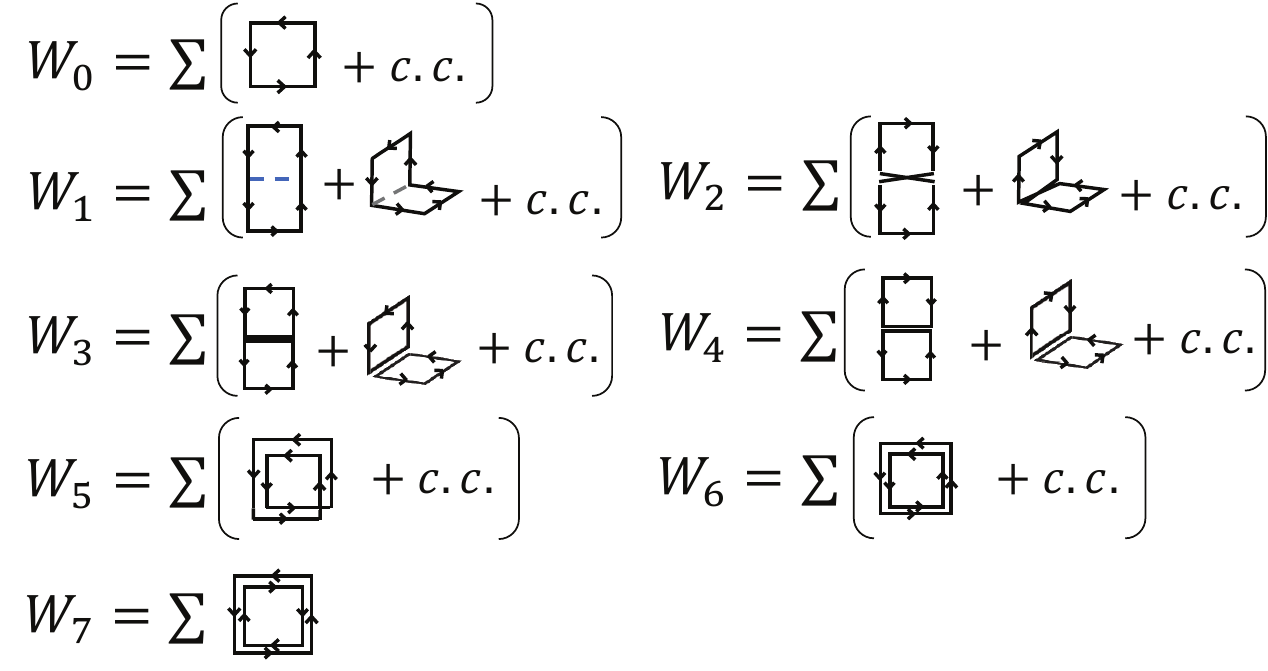}
  \caption{Wilson loops that are relevant in the argument.}
  \label{fig:wilson_loops}
\end{figure}

\begin{figure}[htb]
  \centering
  \includegraphics[width=140mm]{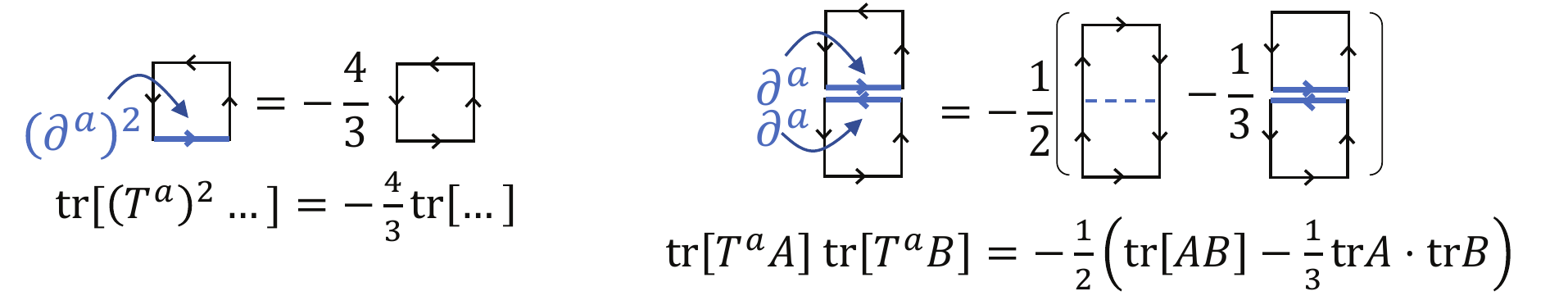}
  \caption{Two representative actions of the differential operators:
    (Left) The plaquette is an eigenfunction of the operator $(\partial^a)^2$. (Right) $\partial^a \cdot \partial^a \cdot$ glues Wilson loops with a trace subtraction.}
  \label{fig:gluing_rules}
\end{figure}

L\"uscher further gave a
way to construct ${\cal F}_{t=1}$
as a $t$-expansion,
which was explicitly demonstrated for the Wilson plaquette action:
\begin{align}
  S_W = -\frac{\beta}{6} W_0,
\end{align}
where $W_0$ is the sum of plaquettes
(see figure~\ref{fig:wilson_loops} for a graphical representation).
The construction begins with expanding $\tilde{S}_t$ as
\begin{align}
  \tilde{S}_t = \sum_{k\geq 0} t^k \tilde{S}^{(k)},
\end{align}
whose convergence radius is proven to be finite
for finite lattice \cite{Luscher:2009eq}.
Plugging the expansion into eq.~\eqref{eq:diff_eq},
we obtain the recurrence relation:
\begin{align}
  &-(\partial^A)^2 \tilde{S}^{(0)} \stackrel{*}{=} S_W, \\
  &-(\partial^A)^2 \tilde{S}^{(k)} \stackrel{*}{=} -(\partial^A S_W)(\partial^A \tilde{S}^{(k-1)}) \quad (k\geq 1).
\end{align}
These relations will be solved in the space of Wilson loops.
Note that the derivative operator $\partial_{x,\mu}^a$
inserts $T^a$ before $U_{x,\mu}$ or
inserts $-T^a$ after $U_{x,\mu}^\dagger$.
The contraction will then be calculated by the completeness relation;
for complex $3\times 3$ matrices $A$ and $B$,
\begin{align}
  {\rm tr}[T^a A T^a B ] &= -\frac{1}{2} \Big({\rm tr}A{\rm tr}B - \frac{1}{3} {\rm tr}AB \Big), \\
  {\rm tr}[T^a A] {\rm tr}[T^a B ] &= -\frac{1}{2} \Big({\rm tr}AB - \frac{1}{3} {\rm tr}A{\rm tr}B \Big).
\end{align}
See figure~\ref{fig:gluing_rules} for examples shown graphically.
Since $W_0$ is an eigenfunction of $-(\partial^A)^2$,
we have for the leading order:
\begin{align}
  \tilde{S}^{(0)} = -\frac{\beta}{32} W_0.
\end{align}
To obtain the next-to-leading order,
we need the Wilson loops $W_0, \cdots, W_7$ shown in figure~\ref{fig:wilson_loops}.
The operator $-(\partial^A)^2$ can be represented in this subspace,
and after inverting the matrix we get
\begin{align}
  \tilde{S}^{(1)} =  \frac{\beta^2}{192}
  \Big(
  -\frac{4}{33}W_1
  +\frac{12}{119}W_2
  +\frac{1}{33}W_3
  -\frac{5}{119}W_4
  +\frac{3}{10}W_5
  -\frac{1}{5}W_6
  +\frac{1}{9}W_7
  \Big).
\end{align}
Since the operation $\partial^A \cdot \partial^A \cdot$
glues the Wilson loops in all possible ways,
the number of relevant Wilson loops
increases for the higher orders in a combinatorial manner.

\subsection{Another look at the map from the space of effective actions}
\label{sec:another_look_eff_action}

The reason for the rapid increase of the involved Wilson loop
can be understood as follows.
First note that, as can be seen in
eq.~\eqref{eq:config_space_flow},
the flow runs from the trivial theory to a nontrivial theory.
Correspondingly, the $t$-expansion around $t$ = 0
is an expansion around the trivial theory.
At small flow times,
we only need to add the plaquettes in the effective action,
and thus the expansion begins with $W_0$.
However, as we evolve the flow,
the effective action goes through the non-trivial theories,
at which the wave-like particle picture should become relevant.
To decrease these modes directly,
we expect that the extended Wilson loops
(presumably summed in certain linear combinations) become relevant.
Therefore, obtaining the exact kernel for the trivializing map is extremely difficult,
and thus we are forced for practical reasons
to choose a finite basis and construct
the efficient flow kernel that decreases the autocorrelations
within the chosen subspace.

For this purpose,
we reexamine the trivializing map
from a different point of view,
namely, from the effective action.
As we demonstrate momentarily,
while the flow in configuration space, eq.~\eqref{eq:config_space_flow},
evolves from the trivial theory to the finite $\beta$ theory,
in the action perspective 
the flow time runs in the opposite direction,
from the finite $\beta$ theory to the trivial theory.
We note that the difference is just the way to see the map,
and, if solved exactly, the two maps coincide.
The difference, however, can occur when we
approximate the map,
and such a functional space point of view
allows us to construct the map
without using the $t$-expansion,
whose example is the Schwinger-Dyson method
we give in section~\ref{sec:schwinger_dyson}.
To distinguish the maps in the two different viewpoints,
we add the superscript $(c)$ to the expressions derived in
the previous configuration space viewpoint.

We define the effective action at time $t$,
$S_t$ (we drop the subscript ${\rm eff}$
for notational simplicity),
by the following recurrence relation:
\begin{align}
  & S_{t=0}(V) \equiv S_W(V), \label{eq:req1}\\
  & S_{t+\epsilon}(V) \equiv S_t({\cal F}_{t,\epsilon}(V)) - \ln \det {\cal F}_{t,\epsilon,*}(V), \label{eq:req2}
\end{align}
where ${\cal F}_{t,\epsilon}$ is again the map that increases $t$
by an amount $\epsilon$, 
and ${\cal F}_{t,\epsilon,*}(V)$ is its Jacobian matrix. 
We here require that ${\cal F}_{t,\epsilon}$ satisfies the relation: 
\begin{align}
  S_{t+\epsilon}(V) \stackrel{*}{=} S_t(V) - \epsilon S_W(V),
  \label{eq:req}
\end{align}
up to an irrelevant constant which we again ignore.
Using eqs.~\eqref{eq:req1} and \eqref{eq:req}, we obtain:
\begin{align}
  S_t(V) \stackrel{*}{=} (1-t) S_W(V), 
  \label{eq:property}
\end{align}
and thus we have the trivializing map at $t=1$.

Note that the composition ordering of ${\cal F}_{t,\epsilon}$ will be
the opposite from eq.~\eqref{eq:composition_ordinary}. 
In fact, from the recurrence formula, we have
\begin{align}
  S_{t=n\epsilon}(V)
  & = S_{(n-1)\epsilon}({\cal F}_{(n-1)\epsilon,\epsilon}(V))
    - \ln \det\mathcal{F}_{(n-1)\epsilon, \epsilon, *}(V) \nonumber\\
  & = S_{(n-2)\epsilon}({\cal F}_{(n-2)\epsilon,\epsilon}\circ{\cal F}_{(n-1)\epsilon,\epsilon}(V)) \nonumber\\
  & \quad \quad - \ln \det\mathcal{F}_{(n-2)\epsilon, \epsilon, *}( {\cal F}_{(n-1)\epsilon,\epsilon}(V))
    - \ln \det\mathcal{F}_{(n-1)\epsilon, \epsilon, *}(V) \nonumber\\
  & = \cdots \nonumber\\
  &= S_0 ({\cal F}_{0,\epsilon}\circ \cdots \circ {\cal F}_{(n-1)\epsilon,\epsilon}(V)) \nonumber\\
  & \quad \quad - \sum_{\ell=0}^{n-1} \ln \det\mathcal{F}_{\ell\epsilon, \epsilon, *}
    ( {\cal F}_{(\ell+1)\epsilon,\epsilon}\circ \cdots \circ {\cal F}_{(n-1)\epsilon,\epsilon}(V) ). 
\end{align}
We identify ${\cal F}_t$ with the composite function: 
\begin{align}
  {\cal F}_t \equiv {\cal F}_{0,\epsilon}\circ \cdots \circ {\cal F}_{(n-1)\epsilon,\epsilon}, 
  \label{eq:composite}
\end{align}
where the Jacobian matrix is given by 
\begin{align}
  {\cal F}_{t,*}(V) \equiv \prod_{\ell=0}^{n-1} \mathcal{F}_{\ell\epsilon, \epsilon, *}
  ( {\cal F}_{(\ell+1)\epsilon,\epsilon}\circ \cdots \circ {\cal F}_{(n-1)\epsilon,\epsilon}(V) ).
  \label{eq:jacobi_composite}
\end{align}
The matrix product is taken in descending order from right to left.
It is easy to see that
$S_t$ is indeed the effective action for the map ${\cal F}_t$:
\begin{align}
  S_t(V) = S_W( {\cal F}_{t}(V) ) - \ln \det {\cal F}_{t,*}(V). 
  \label{eq:S_t_config}
\end{align}

To further rewrite the expression,
we again use the gradient flow ansatz:
\begin{align}
  {\cal F}_{t,\epsilon}(V)_{x,\mu} = e^{ -\epsilon T^a \partial_{x,\mu}^a \tilde{S}_t(V)}V_{x,\mu}.
  \label{eq:kernel_new}
\end{align}
Then the requirement \eqref{eq:req}
gives the functional equation for ${\tilde S}_t$:
\begin{align}
  [- (\partial^A)^2 + (1-t)(\partial^A S_W)\partial^A] \tilde{S}_t \stackrel{*}{=} S_W.
  \label{eq:diff_eq_func}
\end{align}
Comparing eq.~\eqref{eq:diff_eq_func} with \eqref{eq:diff_eq},
we notice that
\begin{align}
  1-t = t^{(c)},
\end{align}
and thus
\begin{align}
  \tilde{S}_t = \tilde{S}^{(c)}_{1-t}
\end{align}
and
\begin{align}
  {\cal F}_{t,\epsilon} = {\cal F}^{(c)}_{1-t,\epsilon}.
  \label{eq:corresp}
\end{align}
From eqs.~\eqref{eq:composite} and \eqref{eq:corresp},
the map ${\cal F}_{t}$ in the functional space viewpoint
can be expressed with ${\cal F}^{(c)}_{t,\epsilon}$ and $1=n\epsilon$ as:
\begin{align}
  {\cal F}_{t=m\epsilon}
  & = {\cal F}_{n\epsilon,\epsilon}^{(c)}\circ \cdots \circ {\cal F}_{(n-m+1)\epsilon,\epsilon}^{(c)}.
    \label{eq:F_with_c}
\end{align}
In particular,
\begin{align}
  &{\cal F}_{1=n\epsilon}
    = {\cal F}_{n\epsilon,\epsilon}^{(c)}\circ \cdots \circ {\cal F}_{\epsilon,\epsilon}^{(c)}, \\
  &{\cal F}_{1=n\epsilon}^{(c)}
    = {\cal F}_{(n-1)\epsilon,\epsilon}^{(c)} \circ \cdots \circ {\cal F}_{0,\epsilon}^{(c)}, 
\end{align}
which should agree in $\epsilon\to 0$.

\begin{figure}[htb]
  \centering
  \includegraphics[width=150mm]{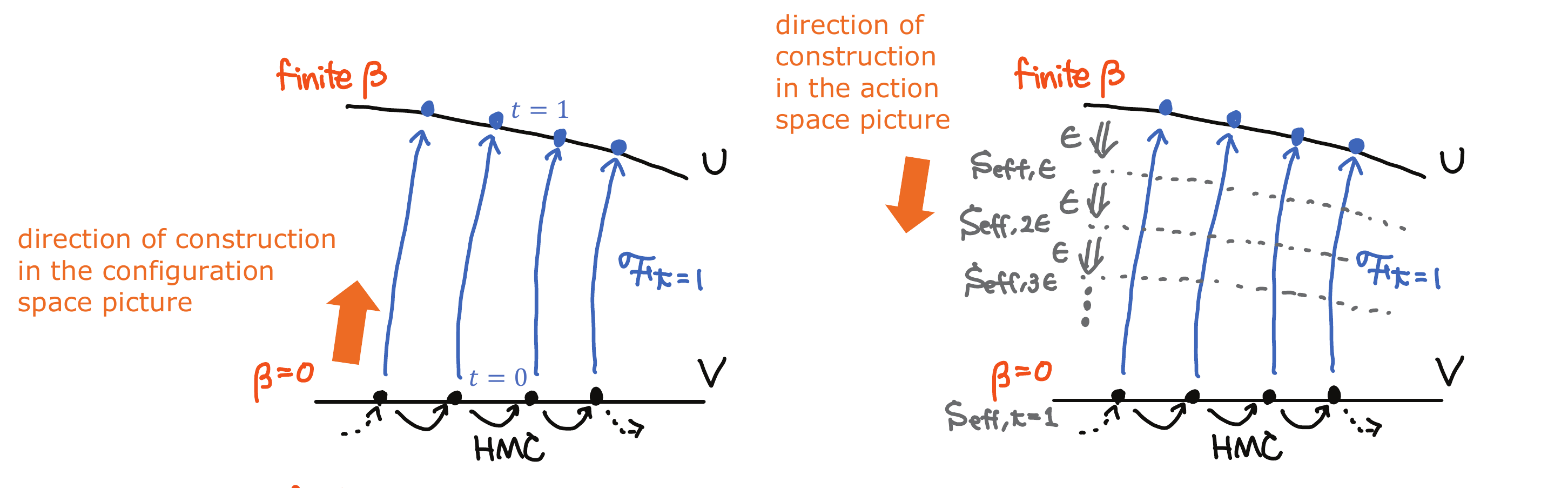}
  \caption{Direction of constructing trivializing maps.}
  \label{fig:direction_construction}
\end{figure}

We therefore have the same trivializing map,
but the direction of the construction is
from finite $\beta$ theory to the trivial theory
(see figure~\ref{fig:direction_construction}).
The complication is that
it is not straightforward to expand in terms of $t$,
as expected from the argument at the beginning of this section.
We will solve eq.~\eqref{eq:diff_eq_func} by picking up a finite basis,
for which we need a way to project the
entire function space to the truncated space.
For this purpose, we use a Schwinger-Dyson equation.

\section{Schwinger-Dyson equation}
\label{sec:schwinger_dyson}

In the Schwinger-Dyson method described below,
we sequentially determine the flow from the finite $\beta$ theory
as explained in section~\ref{sec:another_look_eff_action}.
At each intermediate step,
we determine the effective couplings by a Schwinger-Dyson equation.
The flow will be designed to decrease the couplings from the lattice
expectation values of Wilson loops.

\subsection{Schwinger-Dyson equation for determining coupling constants}
\label{sec:tomography}

In this subsection,
we review the use of a Schwinger-Dyson equation
to determine the couplings in effective actions,
based on refs \cite{Gonzalez-Arroyo:1986kdu,Gonzalez-Arroyo:1987aga}
(see also \cite{Gonzalez-Arroyo:1987mkb,QCD-TARO:1998nbk}).
Suppose that the effective action $S_{\rm eff}$
is expressed as a sum of Wilson loops (and their products):
\begin{align}
  S_{\rm eff} = \sum_j \beta_j W_j.
  \label{eq:wilson_exp}
\end{align}
Then, the coefficients $\beta_{j}$ satisfy
the linear equation:
\begin{align}
  \sum_j\beta_j \langle \partial^AW_j\partial^AW_i \rangle_{S_{\rm eff}}
  = \langle (\partial^A)^2W_i \rangle_{S_{\rm eff}},
  \label{eq:linear_infinite}
\end{align}
where the expectation value $\langle \cdot \rangle_{S_{\rm eff}}$
is taken with respect to the action ${S_{\rm eff}}$.
In fact, under the variation $\delta V$ using $W_i$ in the gradient flow kernel:
\begin{align}
  \delta V_{x,\mu} \equiv -\epsilon \sum_{x,\mu} T^a (\partial_{x,\mu}^aW_i) \, V_{x,\mu},
\end{align}
the path integral is invariant (we vary all the links $(x,\mu)$ at the same time):
\begin{align}
  0 = \delta \int (dV) e^{-S_{\rm eff}(V)}
  = \epsilon \int (dV) e^{-S_{\rm eff}(V)} [-(\partial^A)^2W_i + (\partial^AS_{\rm eff}) \partial^AW_i ].
  \label{eq:sd_orig}
\end{align}
Combining eqs.~\eqref{eq:wilson_exp} and \eqref{eq:sd_orig},
we obtain eq.~\eqref{eq:linear_infinite}.

The linear equation \eqref{eq:linear_infinite}
allows us to obtain the couplings $\beta_j$
from the lattice expectation values.
However, we generically need an infinite number of basis functions
to fully parameterize the action,
and thus $\langle \partial^AW_j\partial^AW_i \rangle_{S_{\rm eff}}$ in eq.~\eqref{eq:linear_infinite}
becomes an infinite-dimensional matrix.
Practically, we cannot calculate an infinite number of
expectation values, and we need to introduce a truncation.
By choosing a finite basis $\{W_{j'}\}$,
where now $j'$ only runs a finite range,
we approximate the action as
\begin{align}
  S_{\rm eff} \approx {\sum_{j'}}' \beta'_{j'} W_{j'} \equiv S_{\rm eff}'.
  \label{eq:trunc_action}
\end{align}
The prime symbols indicate the truncation.
It then turns out that $\beta_j'$ determined by
the finite-dimensional counterpart of eq.~\eqref{eq:linear_infinite}:
\begin{align}
  {\sum_{j'}}'\beta_{j'}' \langle \partial^AW_{j'}\partial^AW_{i'} \rangle_{S_{\rm eff}}
  = \langle (\partial^A)^2W_{i'} \rangle_{S_{\rm eff}},
  \label{eq:linear_finite}
\end{align}
gives the best approximation of $S_{\rm eff}$
in the sense that they minimize the norm
$\Vert S_{\rm eff} - S_{\rm eff}' \Vert_{S_{\rm eff}} $,
where $\Vert \cdot \Vert_{S_{\rm eff}}$ is the force norm:
\begin{align}
  \Vert S \Vert_{S_{\rm eff}}^2 \equiv \langle (\partial^A S)^2 \rangle_{S_{\rm eff}}.
  \label{eq:norm}
\end{align}
In fact, by subtracting eqs.~\eqref{eq:linear_infinite} and \eqref{eq:linear_finite}
for the range of the index $i'$:
\begin{align}
  {\sum_{j'}}' (\beta_{j'}-\beta_{j'}') \langle \partial^AW_{j'}\partial^AW_{i'} \rangle_{S_{\rm eff}} = 0.
  \label{eq:linear_subt}
\end{align}
This is equivalent to the stationary condition:
\begin{align}
  0&=\frac{\partial}{\partial \beta_{i'}'}\Vert S_{\rm eff} - S_{\rm eff}' \Vert^2_{\rm eff} \nonumber\\
   &=\frac{\partial}{\partial \beta_{i'}'}
     \langle [\partial^A(S_{\rm eff} - S_{\rm eff}')]^2\rangle_{\rm eff} \nonumber\\
   &=-2{\sum_{j'}}' (\beta_{j'}-\beta_{j'}') \langle \partial^AW_{j'}\partial^AW_{i'} \rangle_{S_{\rm eff}}.
\end{align}
Since the norm \eqref{eq:norm} is bounded from below but not from above,
we can generically expect the stationary point to be the minimum.
Therefore, this Schwinger-Dyson method
gives us a systematic way to project effective actions
onto a subspace of actions,
where we can measure its goodness from the force fields in numerical calculations.

\subsection{Designing the kernel with the Schwinger-Dyson equation}
\label{sec:design}

We now apply the Schwinger-Dyson equation
in the determination of the approximate trivializing map.
The idea is to reduce the action with $t$ as in eq.~\eqref{eq:req},
replacing the effective action with the approximate one, eq.~\eqref{eq:trunc_action},
so that we can work completely in the functional space spanned by the finite basis.
Below, we write down equations in this finite basis,
and we drop the prime symbols on $j$ for notational simplicity.
We write as ${\cal S}$ the subspace spanned by
the chosen finite set $\{W_j\}$ of Wilson loops and their products.

For the effective action $S_{{\rm eff}, t}$ at a given time $t$,
we apply the Schwinger-Dyson method
to approximate it in ${\cal S}$:
\begin{align}
  S_{{\rm eff},t} \approx S_{{\rm eff},t}' = {\sum_j}'\beta'_{j,t} W_j.
\end{align}
Here the effective couplings $\beta_{j,t}$ are determined by the linear equation:
\begin{align}
  \big\langle
  -(\partial^A)^2W_i + {\sum_j}'\beta_{j,t}' \partial^AW_j \partial^AW_i
  \big\rangle_{S_{{\rm eff},t}} = 0,
  \label{eq:linear_t}
\end{align}
corresponding to eq.~\eqref{eq:linear_finite}.
We again consider the infinitesimal map of the form:
\begin{align}
  {\cal F}_{t,\epsilon}(V)_{x,\mu} = e^{ -\epsilon T^a \partial_{x,\mu}^a \tilde{S}_t(V)}V_{x,\mu},
  \label{eq:each_step}
\end{align}
where this time we construct $\tilde{S}_t$ in the truncated space ${\cal S}$:
\begin{align}
  \tilde{S}_t = {\sum_k}' \gamma_{k,t} W_k.
\end{align}
By taking $d/dt$ in eq.~\eqref{eq:linear_t},
we obtain the linear equation:
\begin{align}
  {\sum_k}'\gamma_{k,t}
  \big\langle
  \partial^BW_k \partial^B
  \big[-(\partial^A)^2W_i + {\sum_j}'\beta_{j,t}' \partial^AW_j \partial^AW_i\big]
  \big\rangle_{S_{{\rm eff},t}} =
  -{\sum_j}'\dot{\beta}_{j,t}'
  \big\langle \partial^AW_j \partial^AW_i \big\rangle_{S_{{\rm eff},t}}.
  \label{eq:gamma_equation}
\end{align}
The left hand side comes from the $t$-dependence of the Boltzmann weight,
and the right hand side from the explicit $t$-dependence of $\beta_{j,t}'$.
Equation~\eqref{eq:gamma_equation}
gives us the coefficients $\gamma_{k,t}$
for a given $\dot{\beta}_{j,t}'$,
i.e., for a given trajectory of $S_{{\rm eff},t}'$ in ${\cal S}$.
We here choose $\dot{\beta}_{j,t}'$ to be:
\begin{align}
  \dot{\beta}_{j,t}' = -\frac{ \beta_{j,t}'}{1-t},
  \label{eq:choice_beta_prime}
\end{align}
so that $\beta_{j,t}' = (1-t)\beta_{j,t=0}'=\beta_{j,t=0}$.\footnote{
  Another choice of $\dot{\beta}_{j,t}'$ leading to the same functional form
  of $\beta_{j,t}$
  is $\dot{\beta}_{j,t}' = -\beta_{j,t=0}$.
  The difference between the two appears in practical calculation where
  we have inexactness due to $O(\epsilon^2)$ terms and the statistical error.
  We choose the form~\eqref{eq:choice_beta_prime} as it
  can take into account of these effects
  from the observed $\beta'_{j,t}$.
  It is, however, uncertain if this is the optimal choice
  though the difference is small compared to
  the truncation effect in choosing ${\cal S}$.
}
After obtaining the coefficients $\gamma_{k,t}$,
and thus the map~\eqref{eq:each_step} at each time step
$t=k\epsilon$, $(k=0,\cdots,n-1)$,
we compose them in the ordering:
\begin{align}
  {\cal F}_t \equiv {\cal F}_{0,\epsilon}\circ \cdots \circ {\cal F}_{(n-1)\epsilon,\epsilon}, 
  \label{eq:composite2}
\end{align}

One of the advantages of designing the trivializing map with the Schwinger-Dyson equation
is that the basis for the flow kernel can be chosen arbitrarily by hand.
Furthermore, it can be applied to the general action of interest,
and the coefficients in the kernel, $\gamma_{j,t}$, are determined by the lattice estimates
of the observables. In this sense,
it contains the non-perturbative information of the theory
and there is no need to make analytic calculations beforehand.
Note also that the
truncation effects and goodness of the map can be measured by the force norm \eqref{eq:norm}.

\begin{figure}[htb]
  \centering
  \includegraphics[width=60mm]{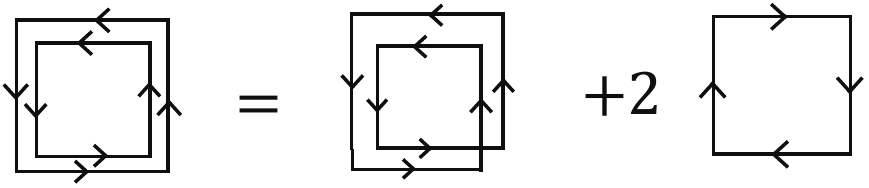}
  \caption{Simplest Mandelstam constraint.}
  \label{fig:mandelstam}
\end{figure}

There is a practical note in the choice of ${\cal S}$.
Since there are linear relations among Wilson loops and their products
called Mandelstam constraints \cite{Mandelstam:1978ed}
(see also \cite{Giles:1981ej,Loll:1992fk,Watson:1993zr}),
we need to choose the basis functions carefully
so that the inversions~\eqref{eq:linear_finite} and \eqref{eq:gamma_equation}
are possible.
The simplest and relevant example is
(see Fig~\ref{fig:mandelstam} for a graphical representation):
\begin{align}
  W_6 = W_5 + 2W_0.
\end{align}
This relation follows from a simple $SU(3)$ identity: For $U \in {\rm SU}(3)$,
\begin{align}
  ({\rm tr}\,U)^2 = {\rm tr}(U^2)+ 2 {\rm tr}\,U^\dagger.
\end{align}
Further relation can be obtained by the Cayley–Hamilton equation:
\begin{align}
  U^3 = ({\rm tr}U)U^2 - \frac{1}{2}\big[({\rm tr}U)^2 - {\rm tr}(U^2)\big] U + 1.
\end{align}

\section{Results}
\label{sec:results}

In this section, we show the evolution of the effective action
and autocorrelation times for the Schwinger-Dyson method.
We use the HMC algorithm with the exact transformed action,
whose detail was given by L\"uscher in \cite{Luscher:2009eq}.
At each step of the approximate trivializing map,
we have the inversion, eq.~\eqref{eq:gamma_equation},
for which we use the numerical derivative with the five-point formula to calculate the matrix
from the flows with $\epsilon=0.0004$.

\subsection{Evolution of the effective action}
\label{sec:control_eff}

\begin{figure}[htb]
  \centering
  \includegraphics[width=90mm]{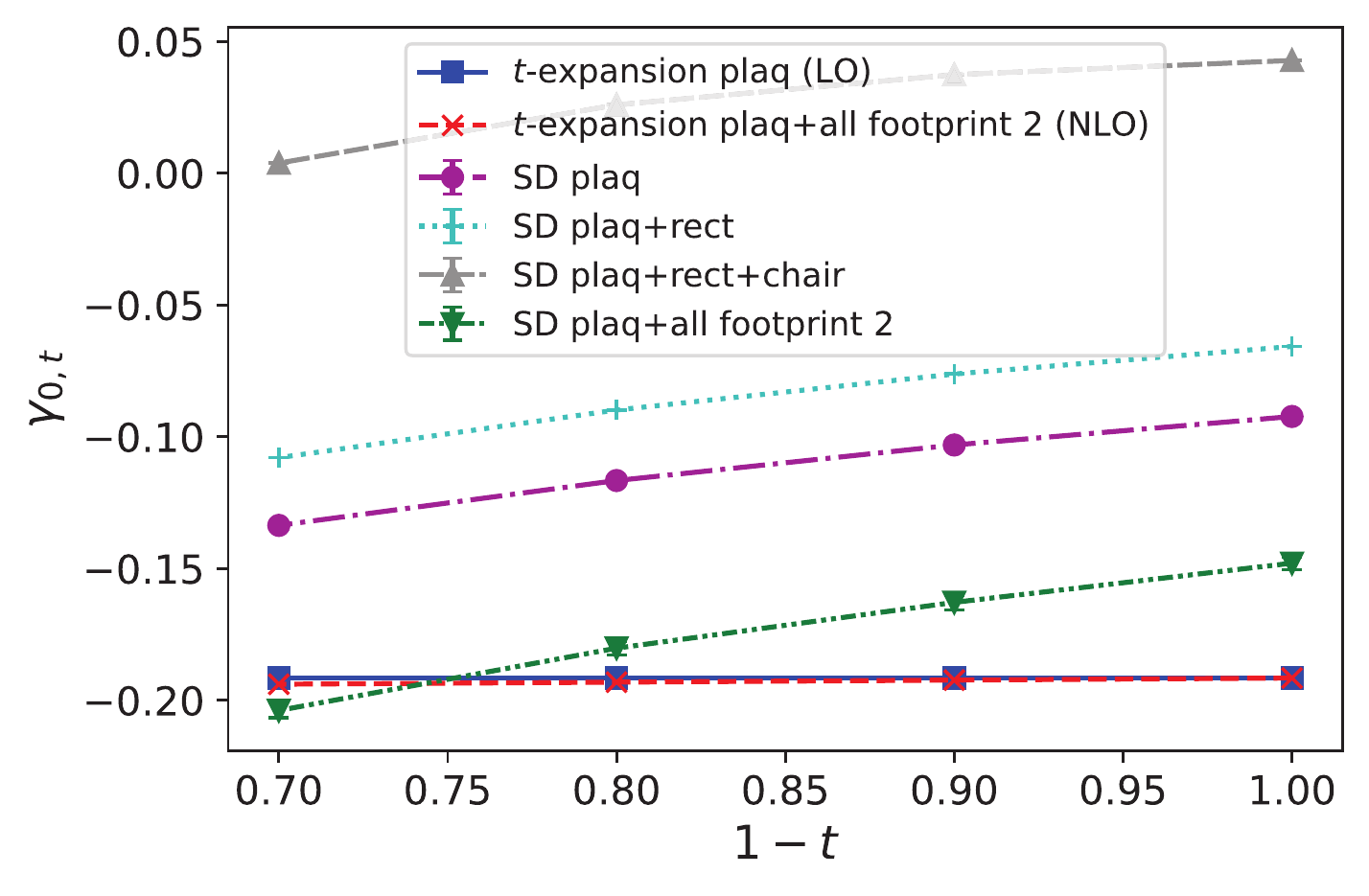}
  \caption{Comparison of $\gamma_{0,t}$ from the $t$-expansion method in the leading order (LO) and the next-to-leading order (NLO) to the Schwinger-Dyson (SD) method with various choices of ${\cal S}$.}
  \label{fig:gamma_0}
\end{figure}

\begin{figure}[htb]
  \centering
  \includegraphics[width=90mm]{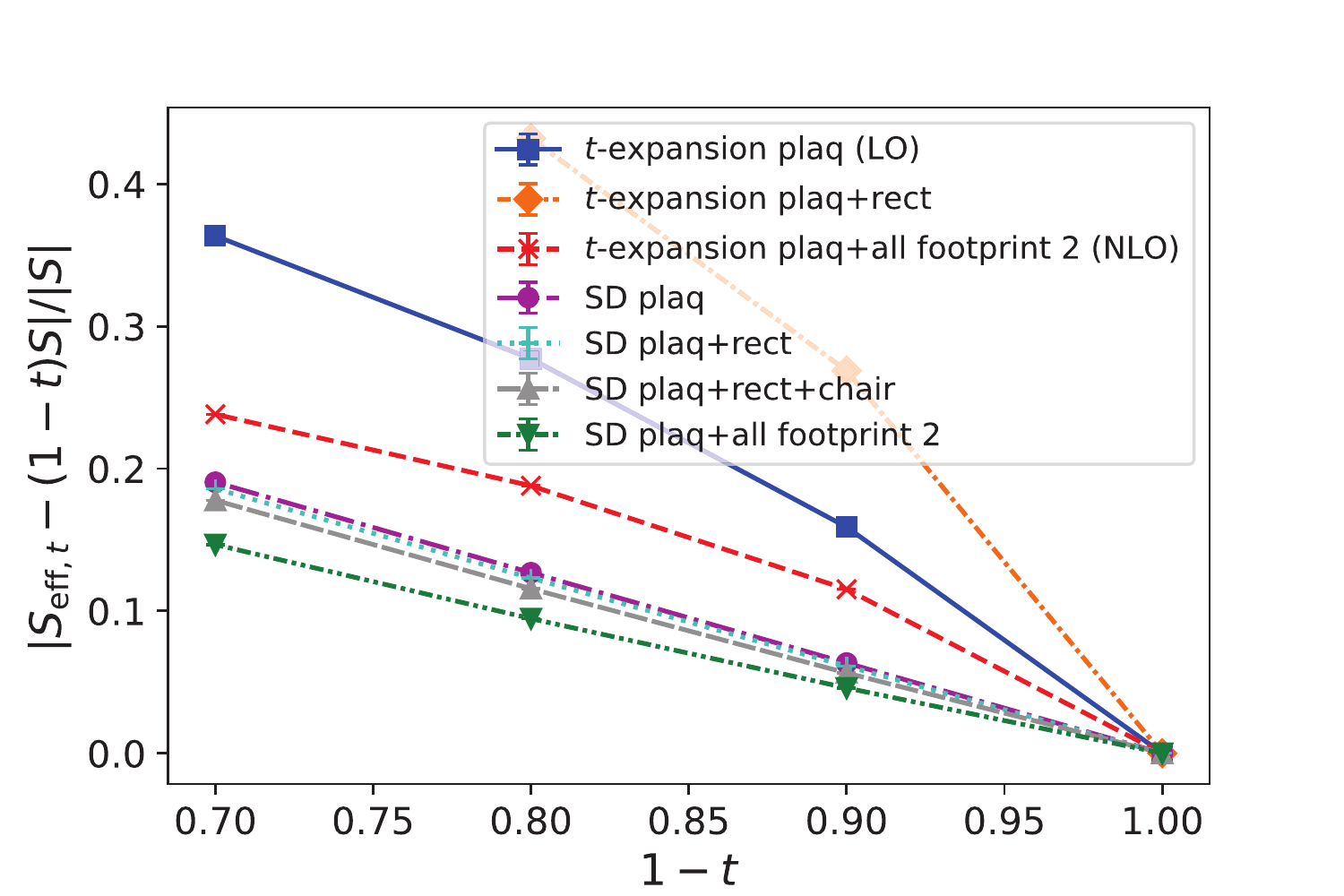}
  \caption{The difference between the effective action $S_{{\rm eff},t}$ and the target action $(1-t)S$.}
  \label{fig:action_norm}
\end{figure}

To compare the effective actions
determined by the Schwinger-Dyson method
to those by the $t$-expansion, we use the Wilson action.
We use $8^4$ lattice with $\beta=6.13$,
which corresponds to $a^{-1}=2.56 {\rm GeV}$ \cite{Ce:2015qha}.
In figure~\ref{fig:gamma_0},
we show one of the determined coefficients, $\gamma_{0,t}$,
which is the coefficient of $W_0$.
We take the step size $\epsilon=0.1$
and consider $t=0.1, \cdots, 0.4$.
We see from the figure that the coefficients $\gamma_{j,t}$ determined by the Schwinger-Dyson method
differ significantly from the $t$-expansion.
It is also notable that
$\gamma_{j,t}$ largely depends on the choice of ${\cal S}$.
In figure~\ref{fig:action_norm},
we plot the norm~\eqref{eq:norm} between $S_{{\rm eff},t}$
and the target action $(1-t)S$ at each time $t$.
We see that, with the Schwinger-Dyson method,
we can have better control of the effective action.
Note that, the norm monotonically decreases
when we enlarge the truncated space ${\cal S}$
though the gain is small compared to
the deviation from the exact trivializing trajectory.
We lastly comment that,
as shown in figure~\ref{fig:action_norm},
naively adding the rectangle term
in the lowest order $t$-expansion formula
(orange dash-dotted line)
does not improve the map, but rather makes it worse;
we need to add all the terms in
the next order expansion
to improve the map in the $t$-expansion method.

\subsection{Autocorrelation}
\label{sec:autocorr}

\begin{figure}[htb]
  \centering
  \includegraphics[width=90mm]{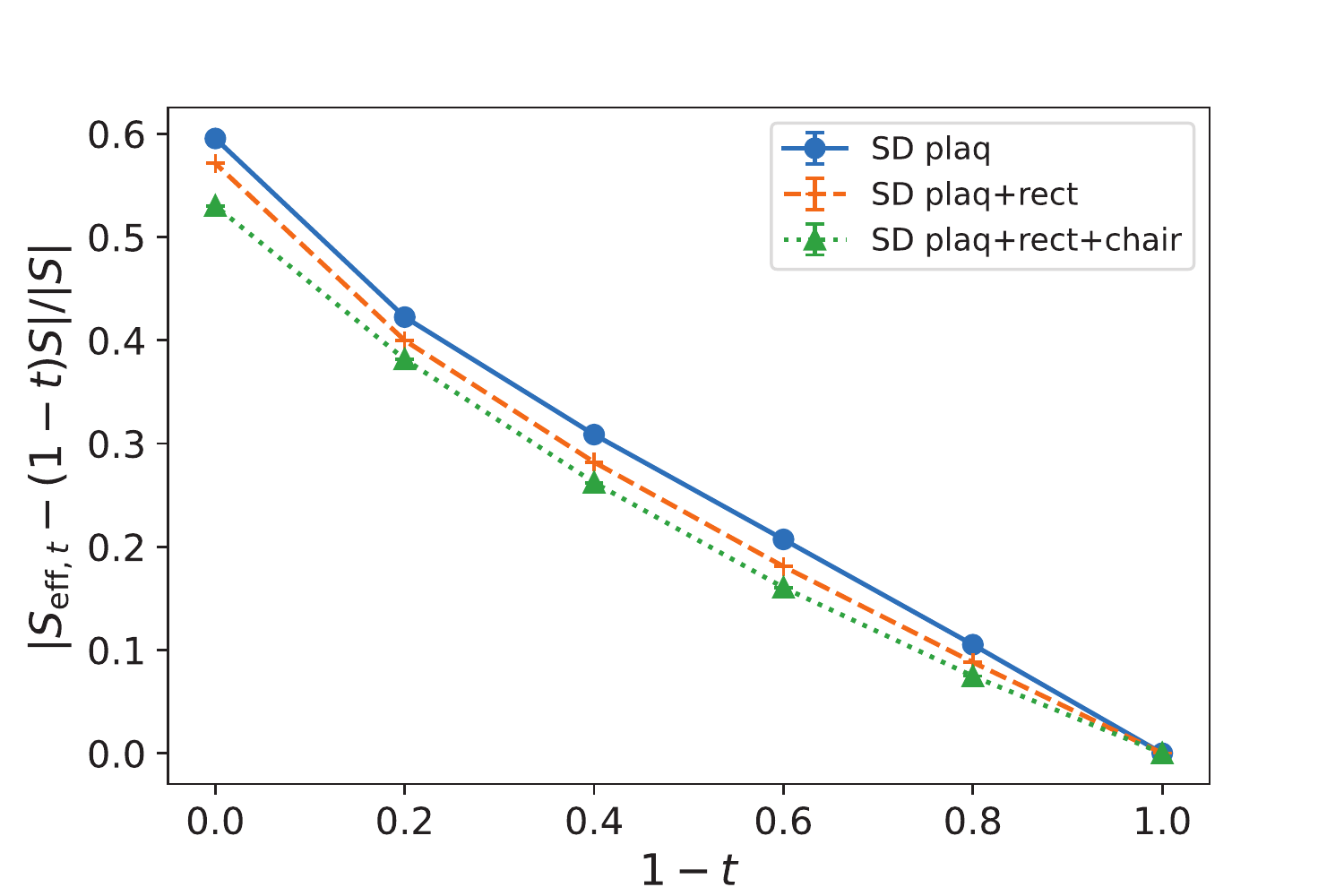}
  \caption{The difference between $S_{{\rm eff},t}$ and the target $(1-t)S$ for the DBW2 case.}
  \label{fig:dbw2_action_diff}
\end{figure}

To study autocorrelation and the efficiency of the algorithm,
we switch to the DBW2 action \cite{QCD-TARO:1996lyt}.
We take $8^3\times 16$ lattice with
$\beta=0.89$, which corresponds to $a^{-1}= 1.49 {\rm GeV}$
\cite{Necco:2003vh}.
In figure~\ref{fig:dbw2_action_diff},
we plot the difference from the target action, 
which is the counterpart of figure~\ref{fig:action_norm} in the DBW2 case.
We choose the step size to be $\epsilon=0.2$
and take $t=0, 0.2, \cdots, 1.0$.
We see that, also for the DBW2 action,
we have better control of the map by enlarging the space ${\cal S}$.

\begin{figure}[htb]
  \centering
  \includegraphics[width=60mm]{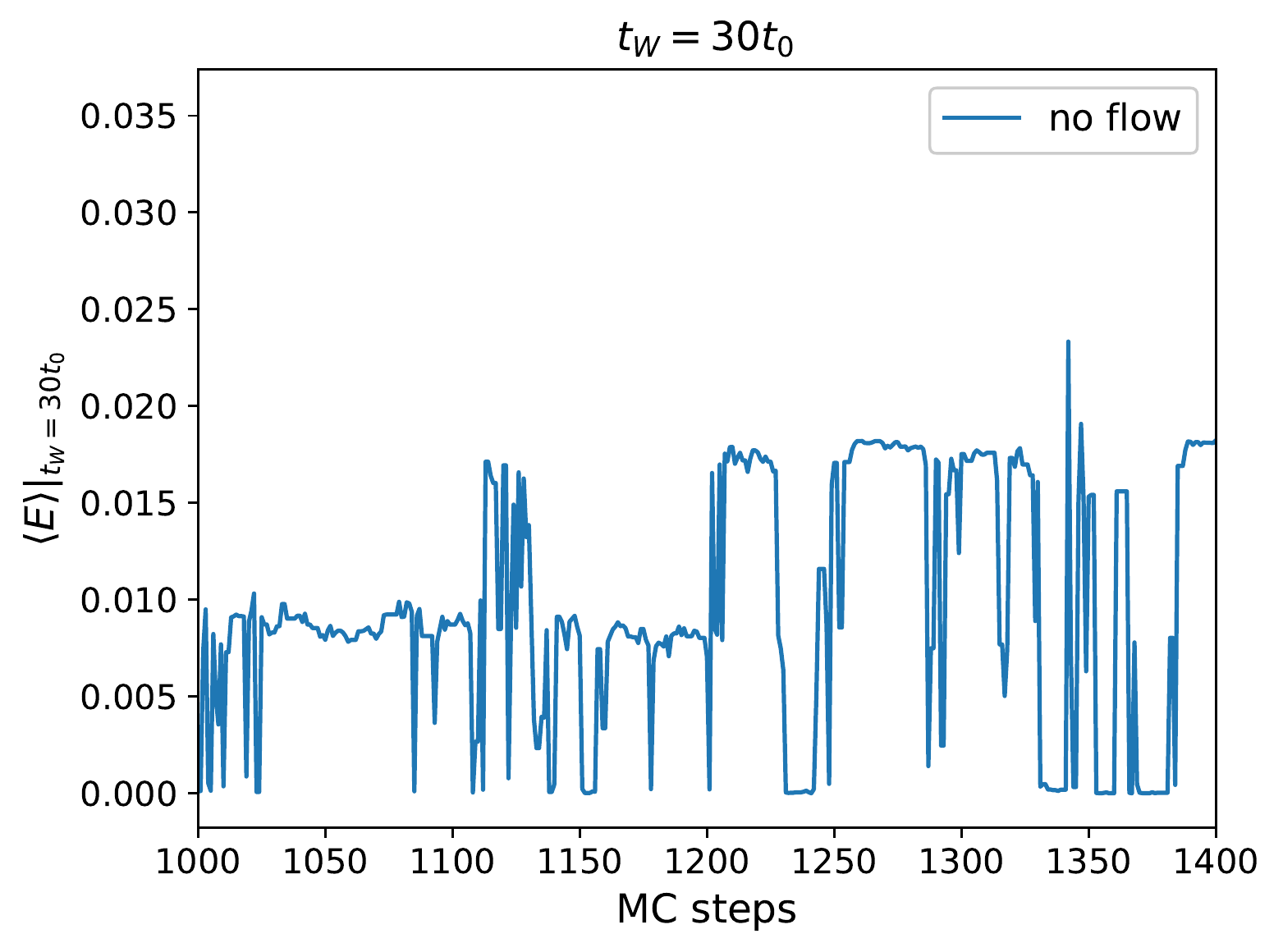}
  \includegraphics[width=60mm]{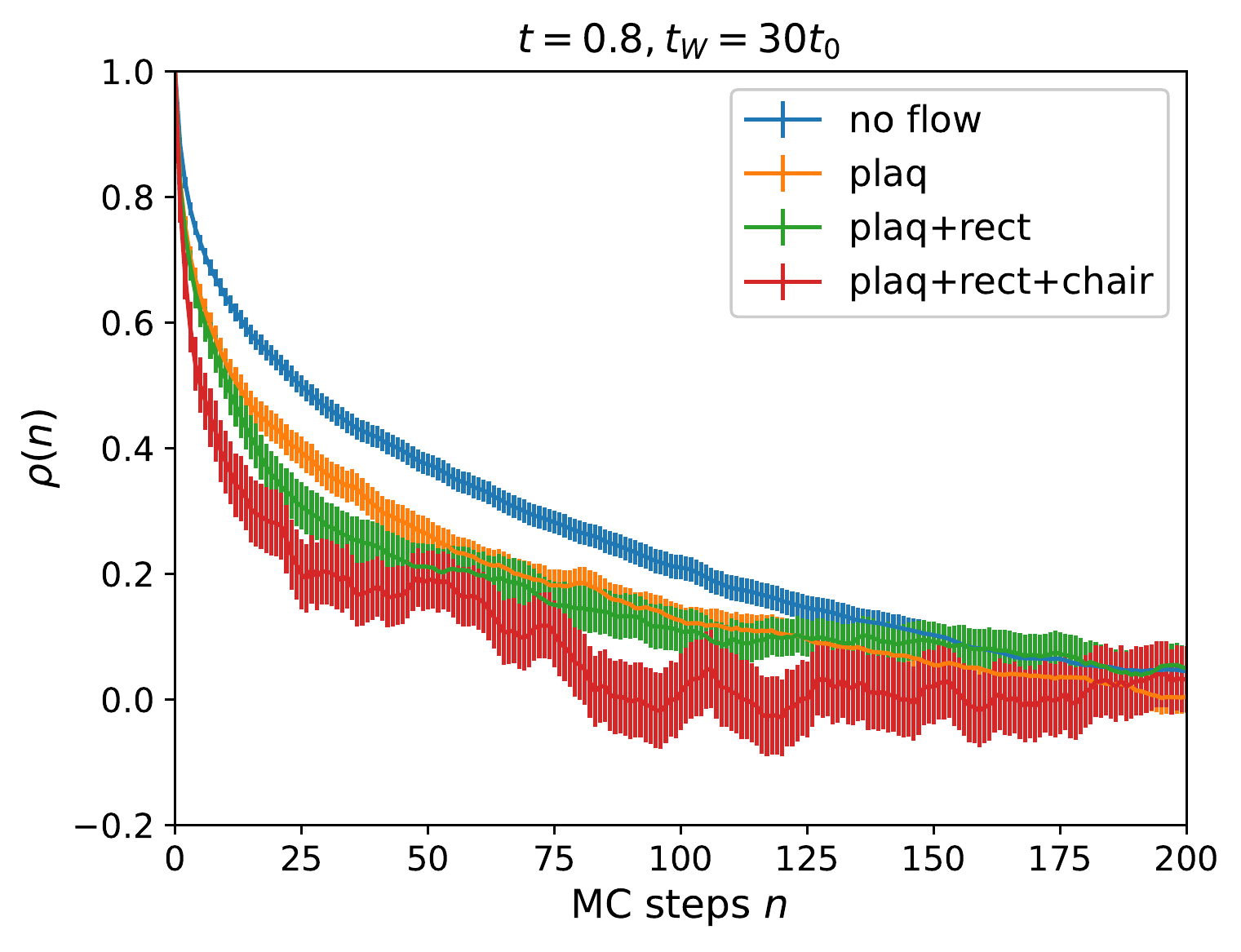}
  \caption{Autocorrelation of the smeared energy density with the Wilson flow time $t_w=30 t_0$.
    (Left) History of the observable generated by the ordinary (not field-transformed) HMC.
    (Right) The normalized autocorrelation function $\rho(n)$ with and without applying the
    approximated trivializing map determined by the Schwinger-Dyson method.
  }
  \label{fig:wilson}
\end{figure}

In figure~\ref{fig:wilson}, we show the history of the
smeared energy density and
the normalized autocorrelation function $\rho_n$
calculated from the data:
\begin{align}
  \rho(n) \equiv C(n) / C(0), \quad
  C(n) \equiv \langle (E_m - \langle E_m \rangle) (E_{m+n} - \langle E_{m+n} \rangle) \rangle.
\end{align}
The smearing is performed with the Wilson flow
with the flow time $t_w = 30t_0$, where $t_0$ is the time scale at which
\cite{Luscher:2010iy}
\begin{align}
  t_w^2 \langle E \rangle = 0.3.
\end{align}
After the smearing time $t_w=30t_0$,
the smeared energy density reflects the instantons
(see the left panel of figure~\ref{fig:wilson}).
The right panel of figure~\ref{fig:wilson}
shows that the configurations decorrelate faster
(in Monte Carlo steps) by including the extended loops.

\begin{figure}[htb]
  \centering
  \includegraphics[width=35mm]{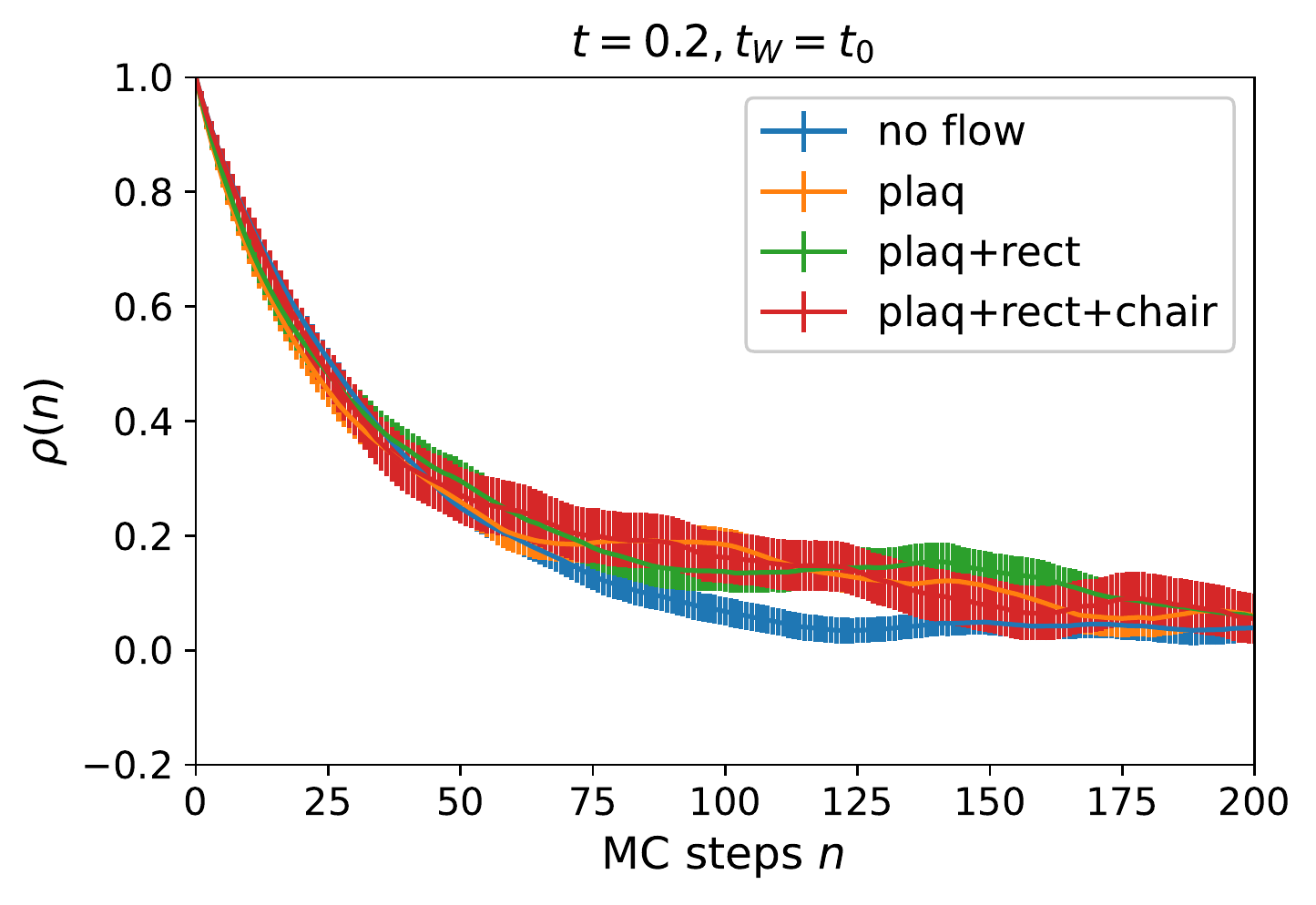}
  \includegraphics[width=35mm]{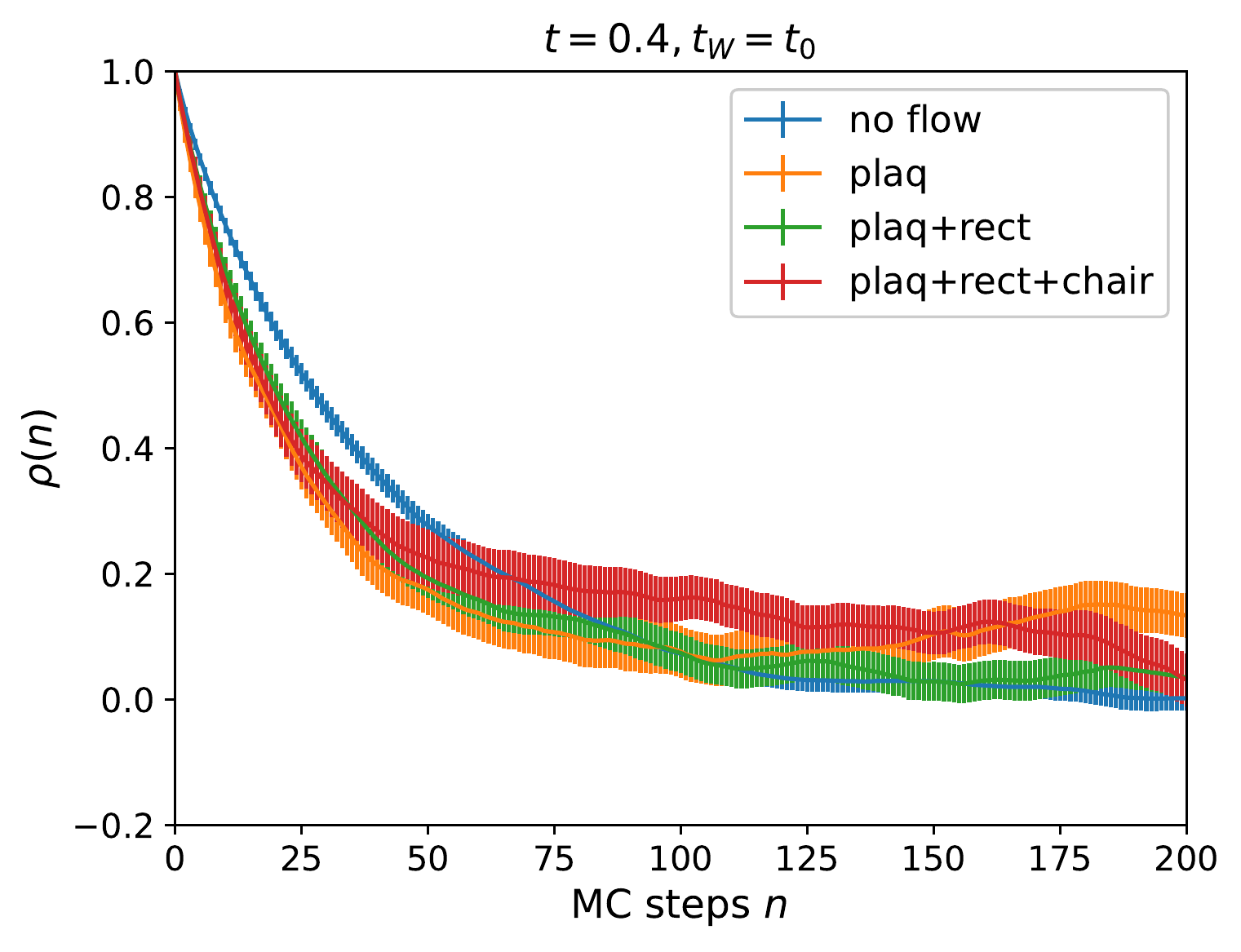}
  \includegraphics[width=35mm]{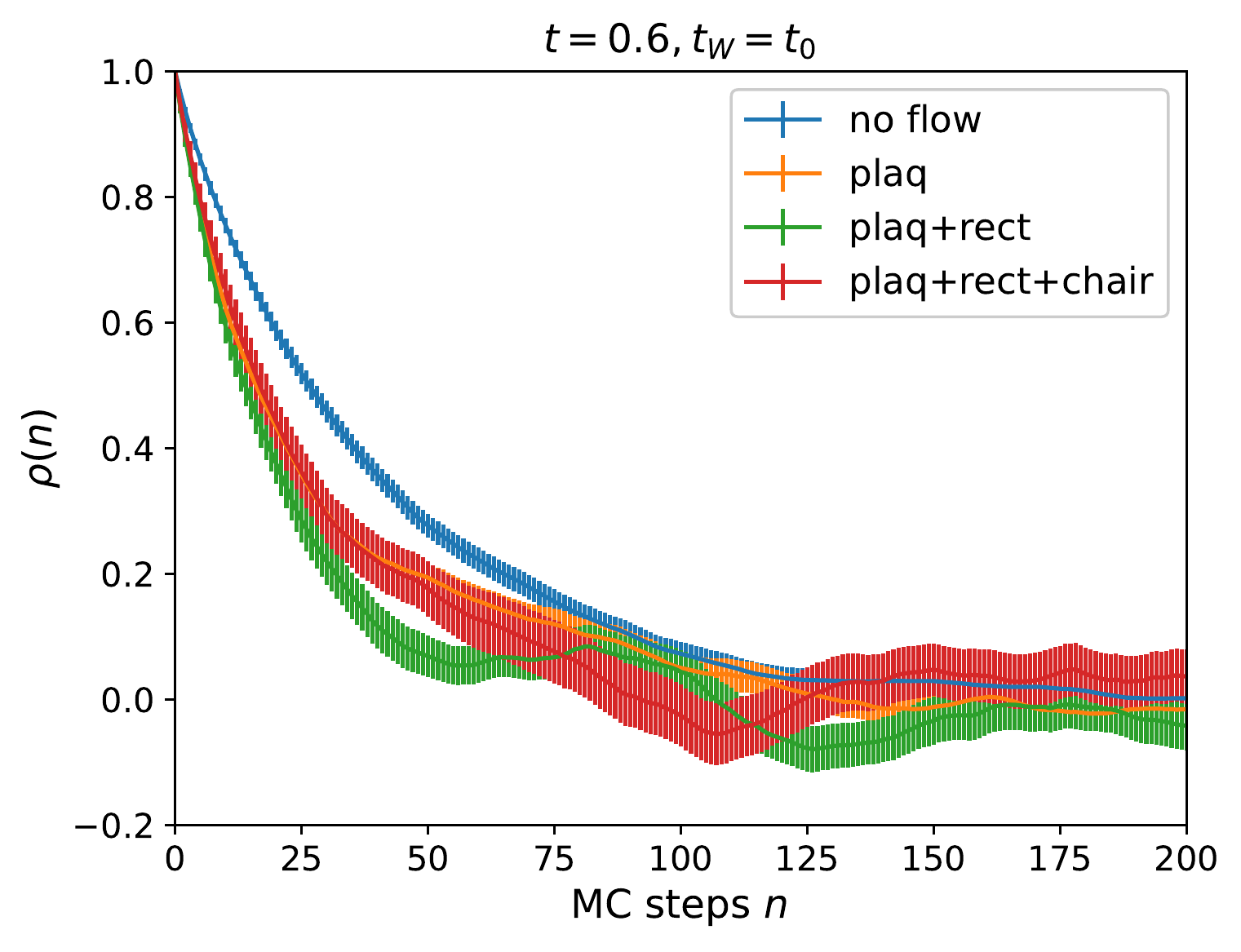}
  \includegraphics[width=35mm]{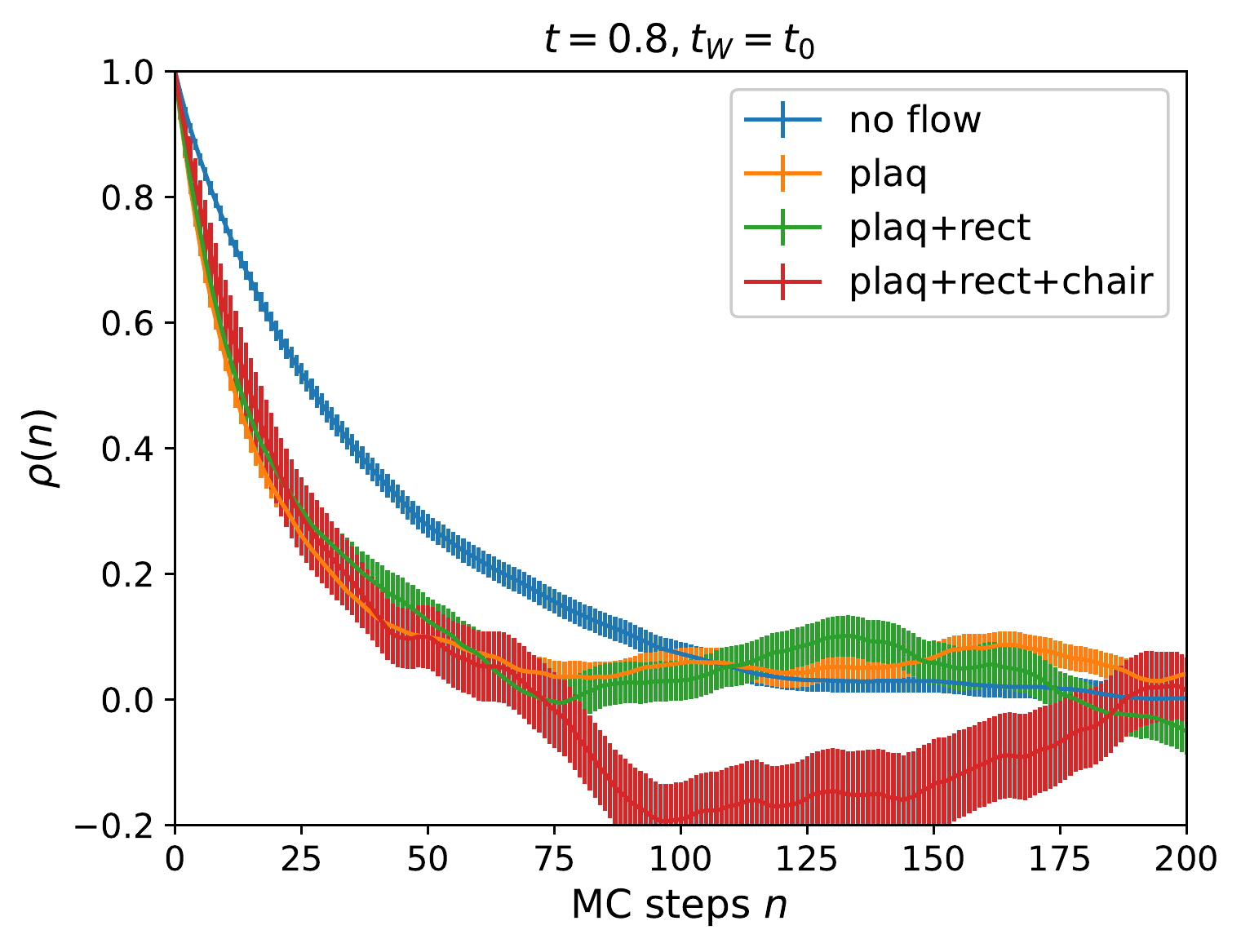} \\
  \includegraphics[width=35mm]{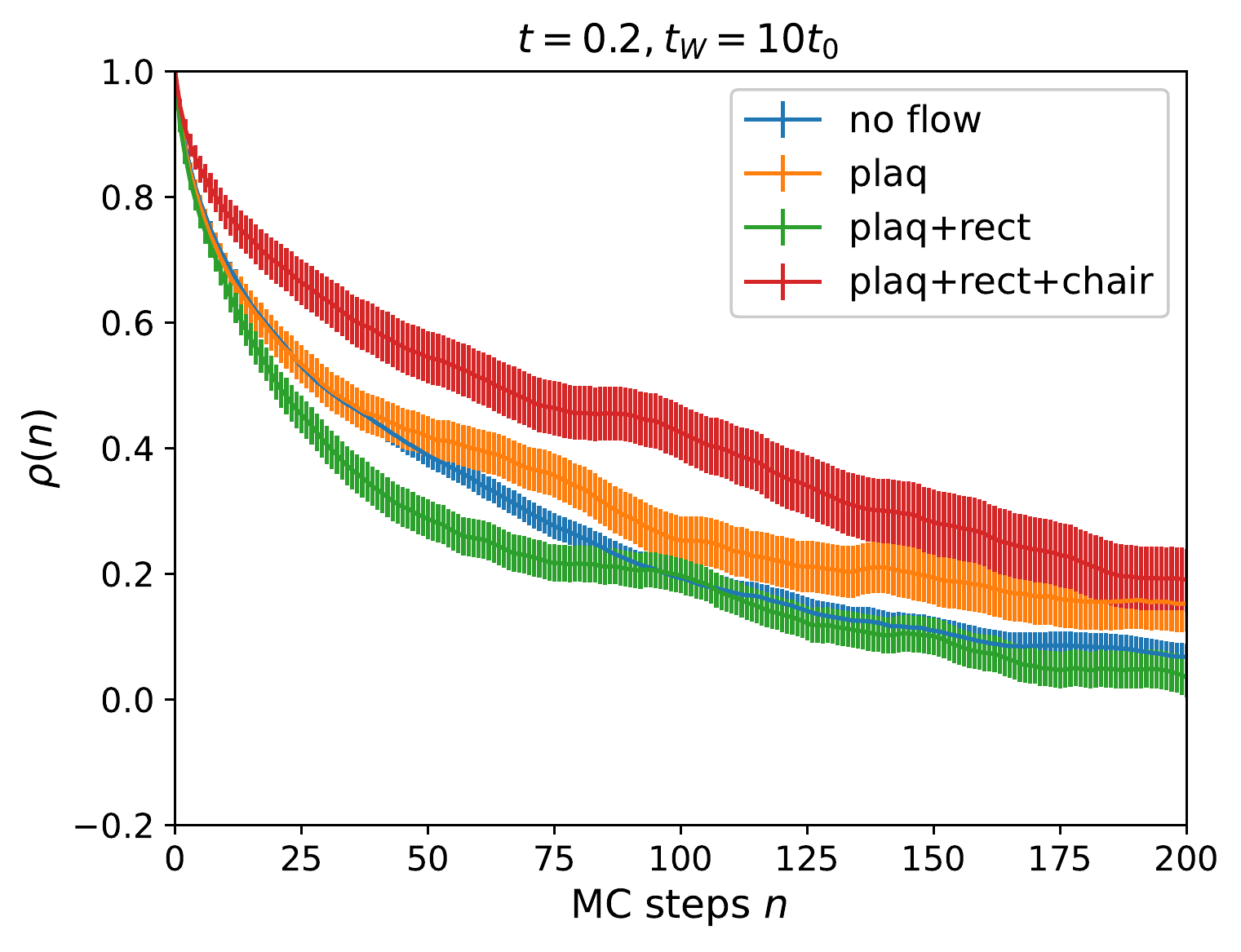}
  \includegraphics[width=35mm]{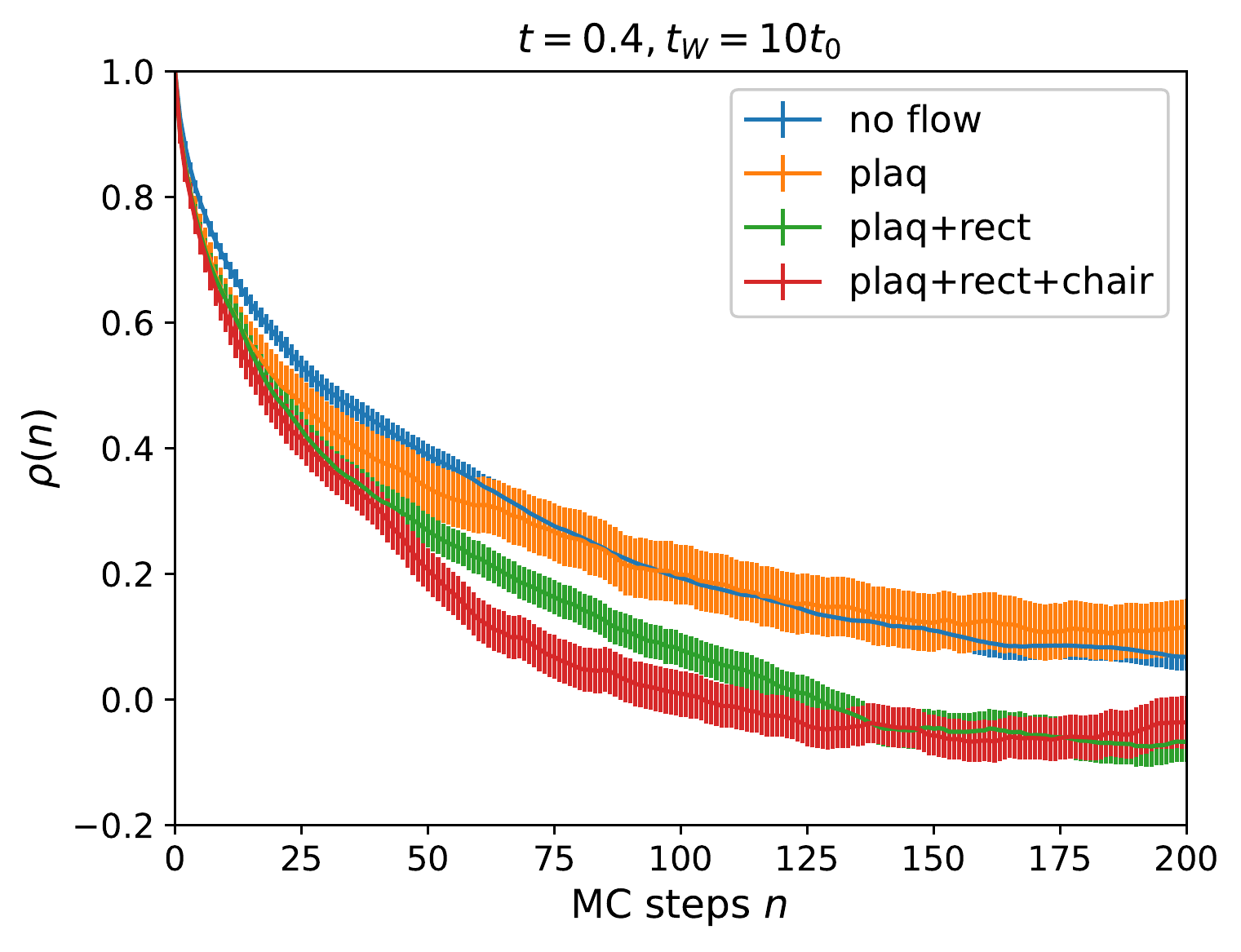}
  \includegraphics[width=35mm]{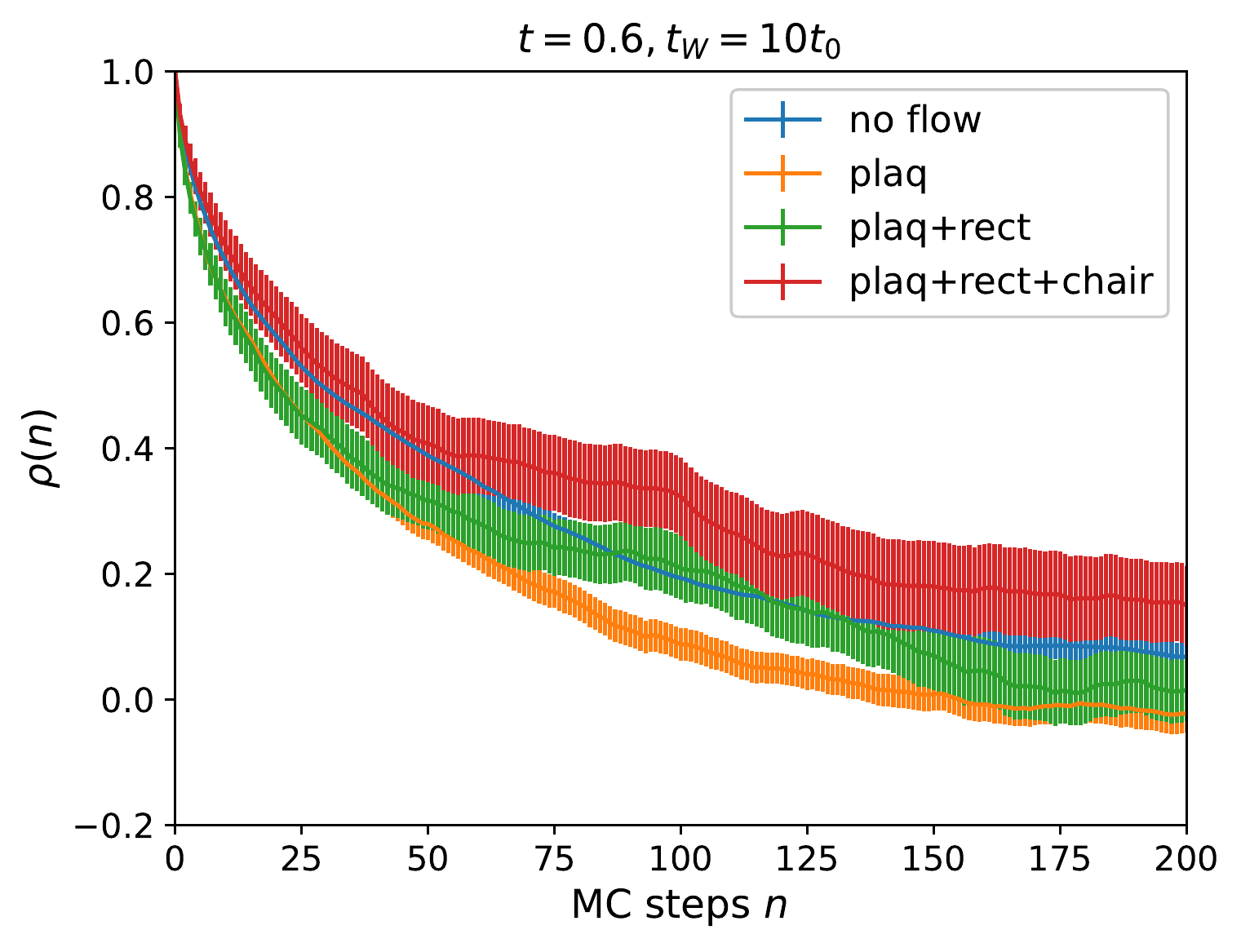}
  \includegraphics[width=35mm]{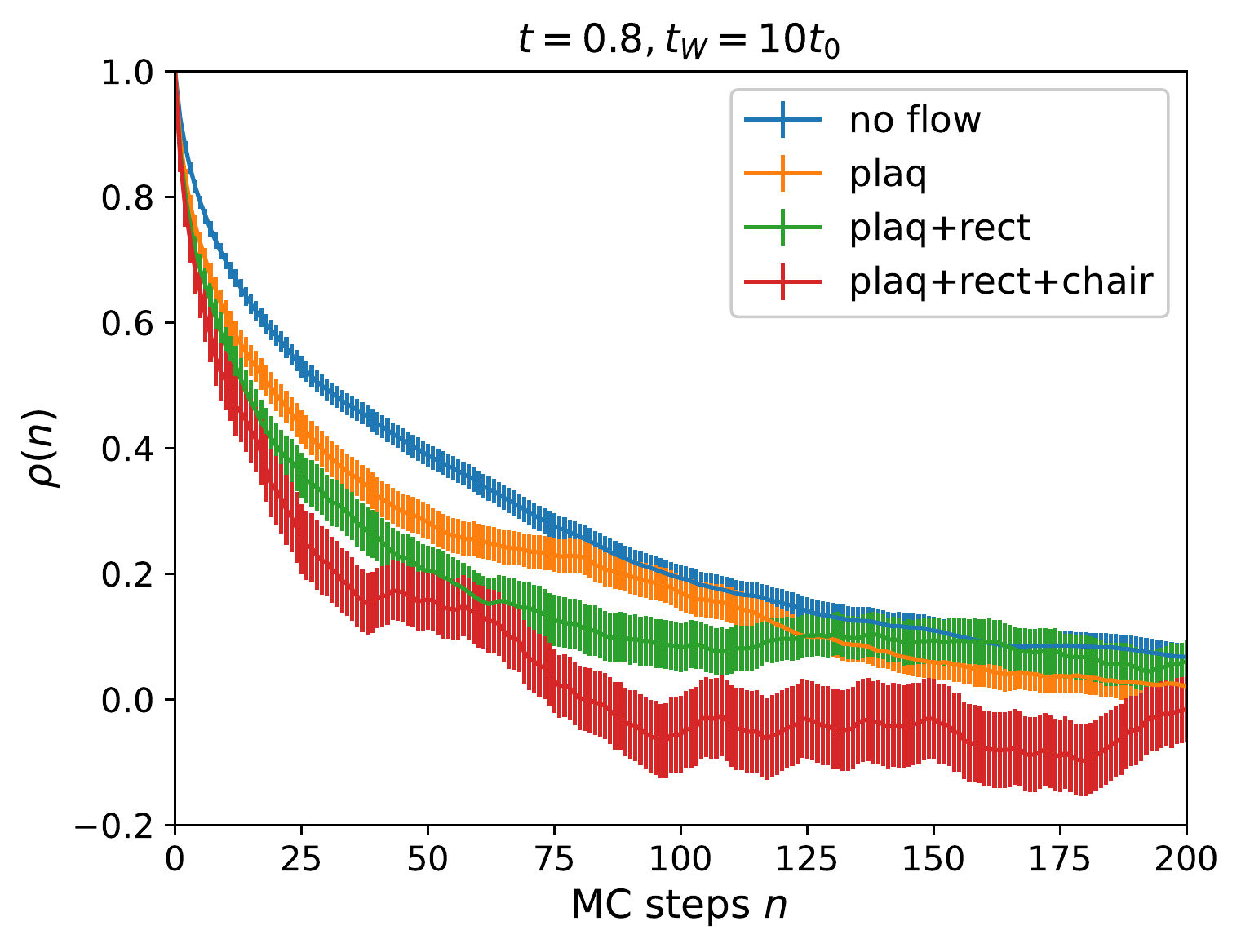} \\
  \includegraphics[width=35mm]{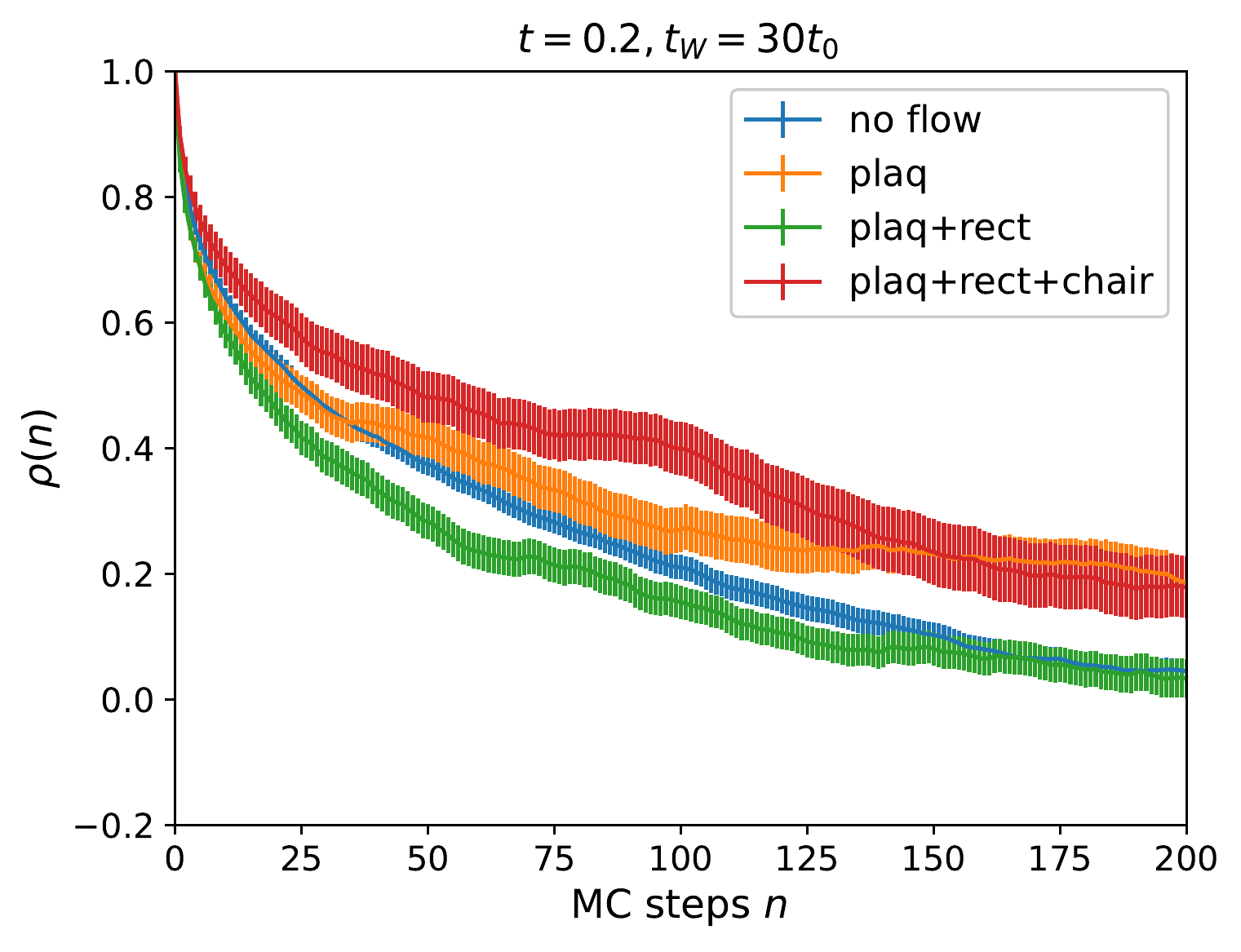}
  \includegraphics[width=35mm]{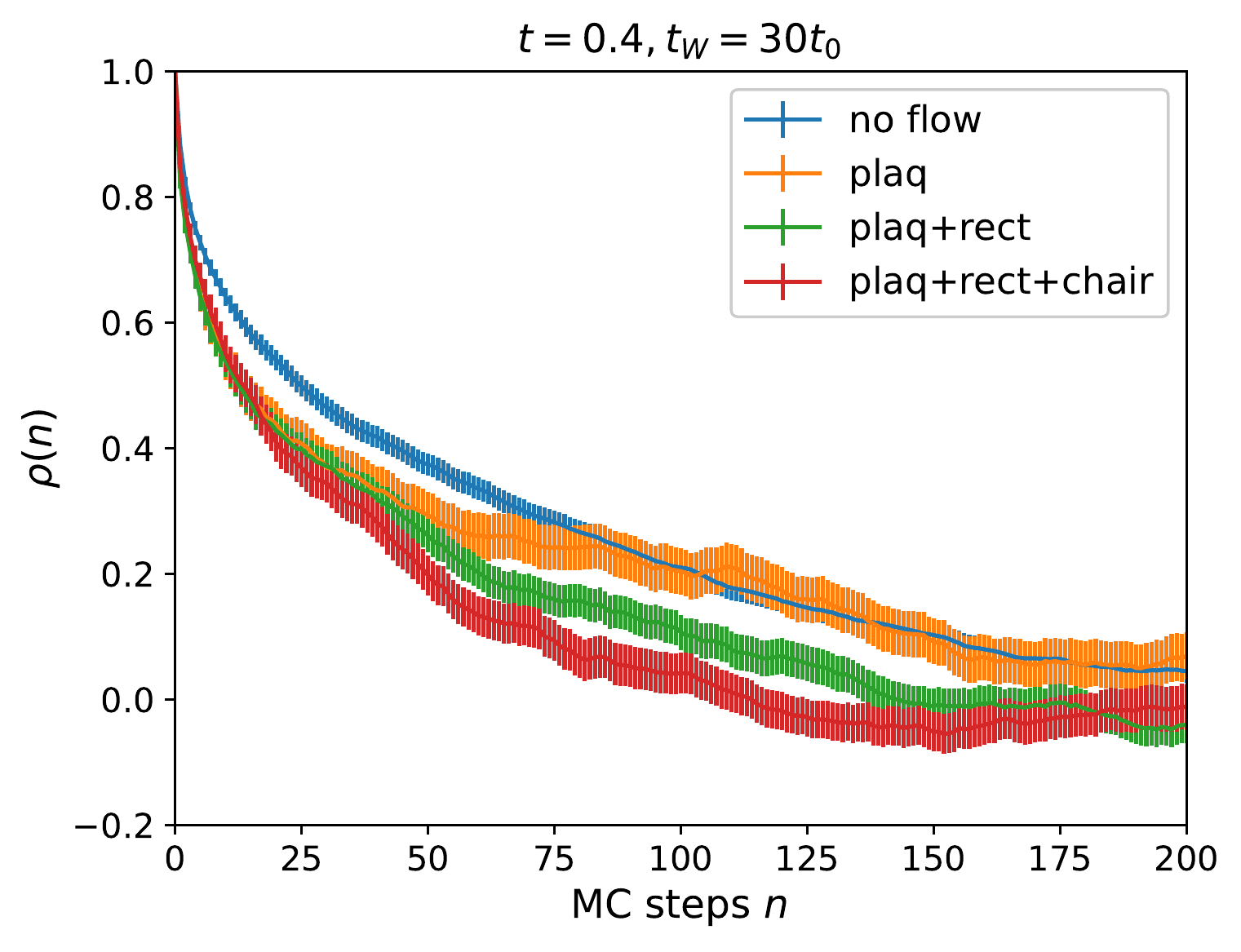}
  \includegraphics[width=35mm]{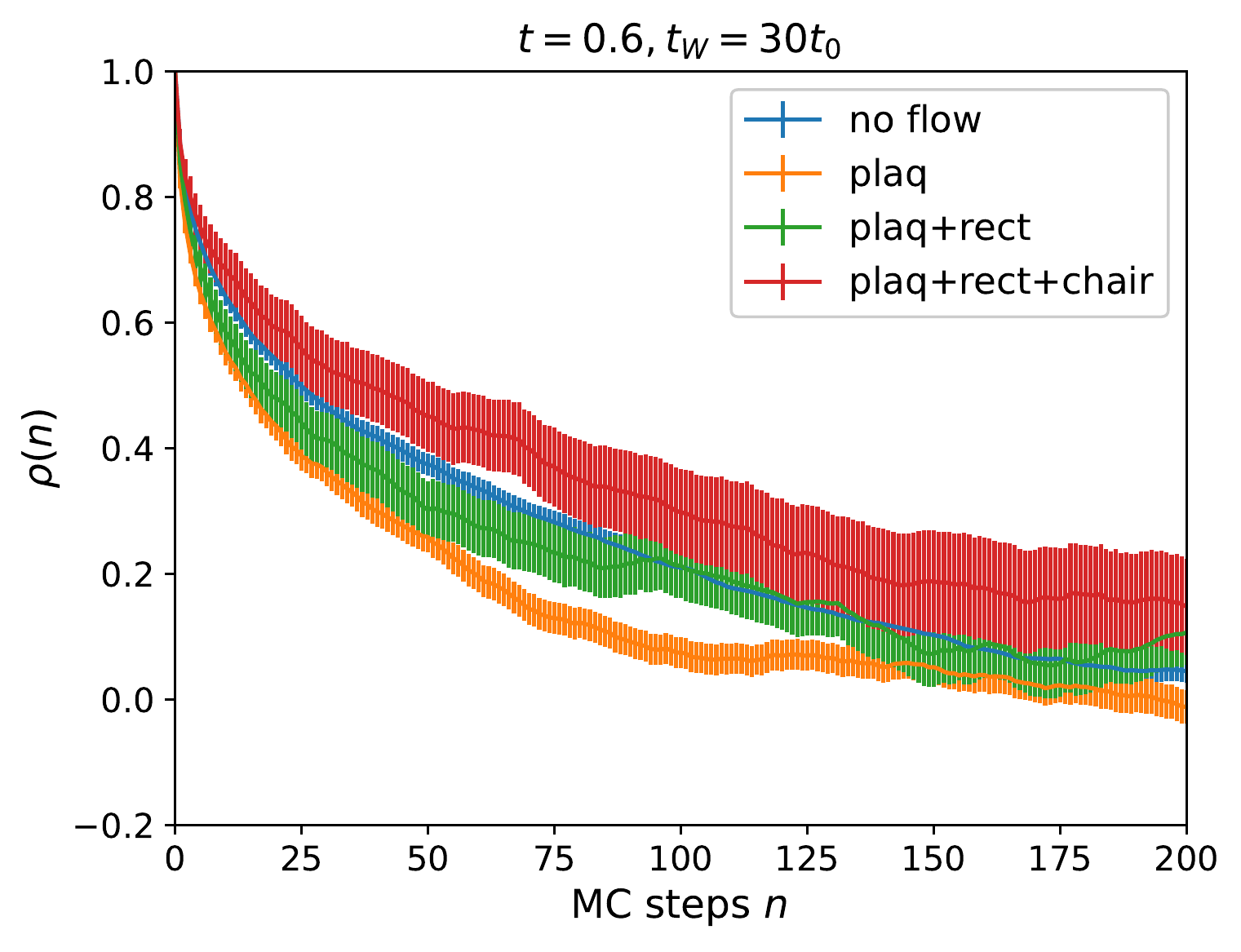}
  \includegraphics[width=35mm]{30t0_4.pdf}
  \caption{The autocorrelation function with various flow time scales}
  \label{fig:various_flow_times}
\end{figure}

However, looking at the observable at the other scales,
we notice that
the autocorrelation is not controlled completely
though we still observe the tendency
of improvement by the extended loops.
In figure~\ref{fig:various_flow_times},
we show the autocorrelation of
the energy densities at smearing times
$t_w=t_0, 10 t_0, 30 t_0$.
As we can see from, e.g., the $t=0.6, t_w = 10t_0$ case,
adding more Wilson loops in $\tilde{S}_t$
does not always reduce the autocorrelation.

\subsection{Algorithmic overhead}
\label{sec:overhead}

\begin{figure}[htb]
  \centering
  \includegraphics[width=60mm]{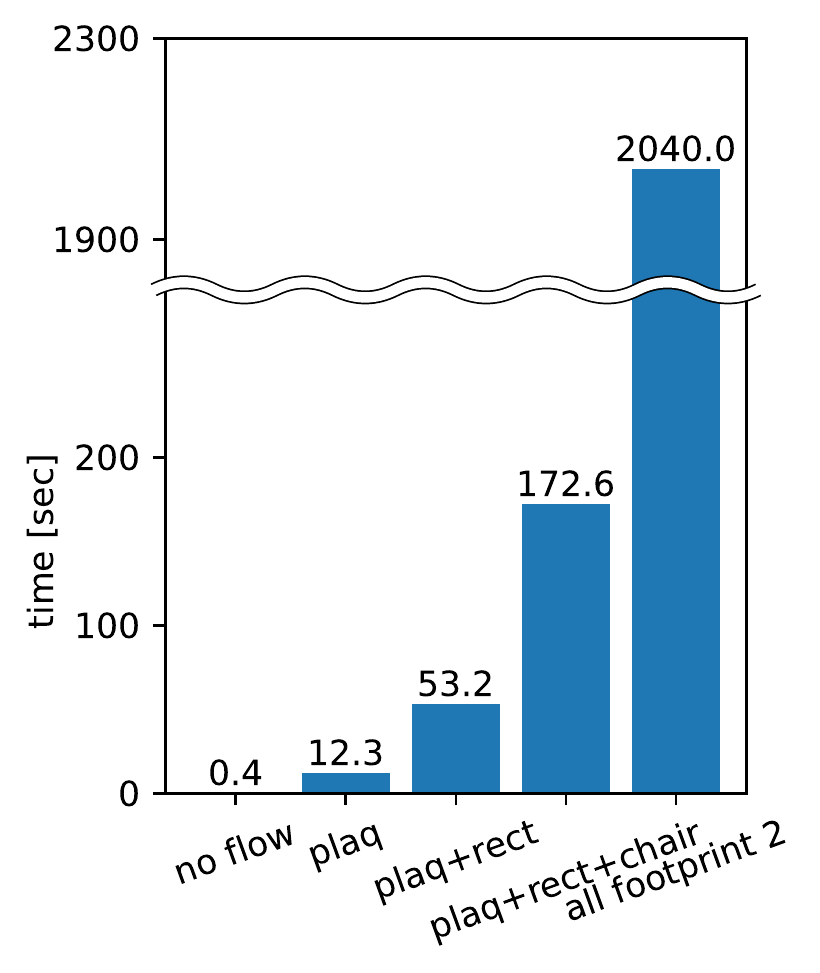}
  \caption{Computation time to generate configuration.}
  \label{fig:calc_time}
\end{figure}

\begin{figure}[htb]
  \centering
  \includegraphics[width=160mm]{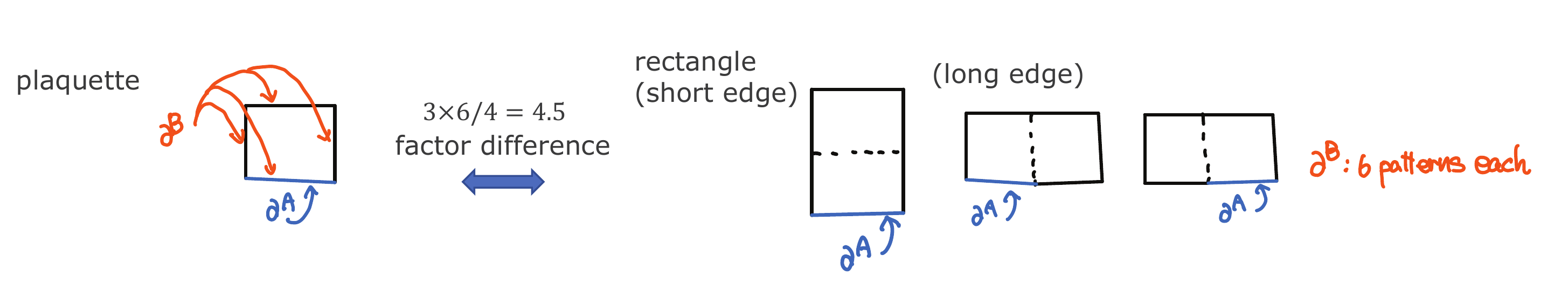}
  \caption{Numerical cost of calculating the Hessian $\partial^A\partial^B {\tilde S}_t$ for plaquette and rectangle.}
  \label{fig:hessian}
\end{figure}

\begin{figure}[htb]
  \centering
  \includegraphics[width=150mm]{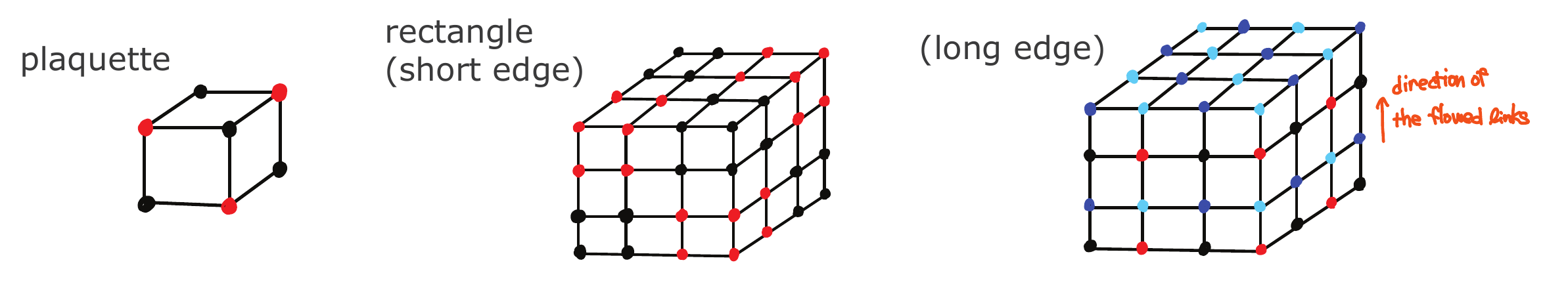}
  \caption{Coloring schemes.}
  \label{fig:color_schemes}
\end{figure}

At this point,
we also mention that
the algorithmic overhead is not negligible
when adding the extended Wilson loops.
Figure~\ref{fig:calc_time} shows the computational cost
for generating a single configuration
with one-step flows with various choices of ${\cal S}$.\footnote{
  This is a one-node calculation fully parallelized with MPI and OpenMP.
}
We see that the cost increases quite rapidly
by adding extended Wilson loops.
The increase can be understood by
the cost of calculating the hessian
$\partial^A \partial^B {\tilde S}_t$,
which is used in the calculation of the Jacobian ${\cal F}_{t,*}$.
Figure~\ref{fig:hessian} pictorially shows the cost difference of calculating
$\partial^A \partial^B W$ for $W=W_0$ (plaquette)
and $W=W_1$ (rectangle).
Here note that, since acting derivatives inserts the generator $T^a$,
each pattern of acting the derivatives
requires evaluation of Wilson loops with different insertions.
By counting the number of ways to choose the links on which $\partial^A$ and $\partial^B$ act,
we have a multiplicative factor of 4.5 in the cost for $W_1$
compared to $W_0$,
which mostly agrees with the cost increase shown in figure~\ref{fig:calc_time}.
We also note that, in the above calculations,
the flow is arranged to make
the Jacobian calculation run in parallel for each link.
Though for the extended shapes we need complicated schemes,
it can be done by dividing the directions of the flowed links
and by appropriately coloring the lattice for
each type of loops \cite{Luscher:2009eq,Boyda:2020hsi}.
Figure~\ref{fig:color_schemes} shows the examples of the coloring schemes.

\section{Discussion}
\label{sec:summary}

In this work,
we proposed a way to design an approximate trivializing map
using a Schwinger-Dyson equation.
The algorithmic advantages of this method are that:
(1) The basis for the flow kernel can be chosen arbitrarily by hand.
(2) It can be applied to the general action of interest.
(3) The coefficients in the kernel are determined by lattice estimates of the observables,
which does not require analytic calculations such as those in the $t$-expansion.
It is also notable that
the truncation effects and goodness of the flow can be measured by the force norm.
We showed that
with the Schwinger-Dyson method, we can have better control of the effective action,
and observed a tendency to decrease the autocorrelation of long-range observables
by adding extended Wilson loops.
However, the decorrelation is not sufficient
compared to the induced overhead.
This indicates that,
though the effective action is getting closer
to the target action,
the large-size Wilson loops that are neglected in the construction
still leave significant contributions.

Indeed, to obtain the exact trivializing map,
in which case we can decrease the autocorrelation
of all the modes of the system,
we need an infinite number of Wilson loops
as reviewed in section~\ref{sec:luscher_construction}.
However, adding more and more Wilson loop basis
is not realistic because of the increasing algorithmic
overhead shown in subsection~\ref{sec:overhead}.
Since it is likely that we need to restrict ourselves to
a few small Wilson loops,
one possible strategy may be to
be more specific to a particular slow observable,
e.g., the topological charge.
For example, as is well known,
the instanton potential with respect to its radius
can be different for the same lattice spacing
but with different gauge action.
We thus may be able to change the instanton potential
to stimulate tunneling
with the small number of basis functions in ${\cal S}$
(cf.
\cite{Smith:1998wt,DeGrand:1997de,
  deForcrand:1997ut,Hasenfratz:1998qk}).
Studies along these lines are in progress and will be reported elsewhere.

\section{Acknowledgments}
\label{sec:ack}

The authors thank Norman Christ,
William Detmold, Sam Foreman, Xiao-Yong Jin,
Gurtej Kanwar, James Osborn, and
Phiala Shanahan for valuable discussions.
Computation was performed in
RIKEN HOKUSAI, Univ of Tokyo Oakforest-PACS.
We have also used a USQCD facility at BNL (KNL),
which is funded by the Office of Science of the U.S. Department of Energy.
AT is supported by JSPS KAKENHI Grant Numbers
JP20K14479, JP22H05112, JP22H05111, and JP22K03539.
NM is supported by JP22H01222 and the Special Postdoctoral Researchers Program of RIKEN.



\end{document}